\documentclass[12pt]{article}

\usepackage{graphicx}
\usepackage{float}
\usepackage{cite}
\usepackage{amsfonts}
\usepackage{amssymb}
\usepackage{amsmath}
\usepackage{relsize}
\usepackage[compatibility=false]{caption}
\usepackage{subcaption}

\def\gtwid{\mathrel{\raise.3ex\hbox{$>$\kern-.75em\lower1ex\hbox{$\sim$}}}}
\def\ltwid{\mathrel{\raise.3ex\hbox{$<$\kern-.75em\lower1ex\hbox{$\sim$}}}}
\def\square{\kern1pt\vbox{\hrule height 1.2pt\hbox{\vrule width 1.2pt\hskip 3pt
   \vbox{\vskip 6pt}\hskip 3pt\vrule width 0.6pt}\hrule height 0.6pt}\kern1pt}

\begin{document}

\begin{titlepage}

\begin{flushright}
UFIFT-QG-23-02
\end{flushright}

\vskip 3cm

\begin{center}
{\bf Coincident Massless, Minimally Coupled Scalar Correlators on General Cosmological Backgrounds}
\end{center}

\vskip 1cm

\begin{center}
E. Kasdagli$^{*}$, M. Ulloa$^{\dagger}$ and R. P. Woodard$^{\ddagger}$
\end{center}

\begin{center}
\it{Department of Physics, University of Florida,\\
Gainesville, FL 32611, UNITED STATES}
\end{center}

\vspace{1.5cm}

\begin{center}
ABSTRACT
\end{center}
The coincidence limits of the massless, minimally coupled scalar propagator
and its first two derivatives have great relevance for the project of summing
up the leading logarithms induced by loops of inflationary gravitons. We use
dimensional regularization to derive good analytic approximations for the 
three quantities on a general cosmological background geometry which underwent 
inflation.

\begin{flushleft}
PACS numbers: 04.50.Kd, 95.35.+d, 98.62.-g
\end{flushleft}

\vspace{3cm}

\begin{flushleft}
$^{*}$ e-mail: kasdaglie@ufl.edu \\
$^{\dagger}$ e-mail: m.ulloacalzonzin@ufl.edu \\
$^{\ddagger}$ e-mail: woodard@phys.ufl.edu
\end{flushleft}

\end{titlepage}

\section{Introduction}

We are interested in quantum field theory on a homogeneous, isotropic
and spatially flat geometry with scale factor $a(t)$, Hubble parameter
$H(t)$, and first slow roll parameter $\epsilon(t)$, 
\begin{equation}
ds^2 = -dt^2 + a^2(t) d\vec{x} \!\cdot\! d\vec{x} \qquad , \qquad H(t) 
\equiv \frac{\dot{a}}{a} \quad , \quad \epsilon(t) \equiv 
-\frac{\dot{H}}{H^2} \; . \label{geometry}
\end{equation}
The dynamical system we study is the massless, minimally coupled scalar,
\begin{equation}
\mathcal{L} = -\frac12 \partial_{\mu} \Phi \partial_{\nu} \Phi
g^{\mu\nu} \sqrt{-g} \; . \label{MMCSL}
\end{equation} 
This system is of great interest because it includes the free inflaton, 
as well as dynamical gravitons \cite{Lifshitz:1945du}. What we specifically
seek are coincidence limits of the scalar propagator $i\Delta(x;x')$ and 
its first two derivatives,
\begin{equation}
A(t) \equiv i\Delta(x;x')\Bigl\vert_{x' = x} , 
B_{\mu}(t) \equiv \partial_{\mu} i\Delta(x;x')\Bigl\vert_{x'=x} ,
C_{\mu\nu}(t) \equiv \partial_{\mu} \partial_{\nu}' i\Delta(x;x') 
\Bigl\vert_{x'=x} . \label{threeDs}
\end{equation}
These correlators are crucial in deriving the curvature-dependent effective 
potentials \cite{Miao:2021gic} needed to apply Starobinsky's stochastic 
formalism \cite{Starobinsky:1986fx,Starobinsky:1994bd} to the task of 
re-summing the large temporal and spatial logarithms induced by loops of 
inflationary gravitons \cite{Miao:2006gj,Glavan:2013jca,Wang:2014tza,
Park:2015kua,Tan:2021lza,Glavan:2021adm,Tan:2022xpn}.

To better motivate this study, note first that graviton loop corrections
on de Sitter background typically involve logarithms of the scale factor
$a$. For example, the Coulomb potential of a stationary charge $Q$ 
becomes \cite{Glavan:2013jca},
\begin{equation}
\Phi(t,r) = \frac{Q}{4\pi a r} \Biggl\{1 + \frac{2 G}{3\pi a^2 r^2}
+ \frac{2 G H^2}{\pi} \ln(a H r) + O(G^2) \Biggr\} \; , 
\end{equation}
where $G$ is Newton's constant. Similarly, a single graviton loop enhances
the electric field strength of a plane wave photon to \cite{Wang:2014tza},
\begin{equation}
F^{0i}(t,\vec{x}) = F^{0i}_0(t,\vec{x}) \Biggl\{ 1 + \frac{2 G H^2}{\pi} 
\ln(a) + O(G^2) \Biggr\} \; ,
\end{equation}
where $F^{0i}_0(t,\vec{x})$ is the tree order field strength. The 
analogous results for the mode function $u(t,k)$ of a plane wave graviton
\cite{Tan:2021lza} and the Newtonian potential of a stationary mass $M$
\cite{Tan:2022xpn} are,
\begin{eqnarray}
u(t,k) &\!\!\! = \!\!\!& u_0(t,k) \Biggl\{ 1 + \frac{16 \pi G H^2}{3\pi}
\ln^2(a) + O(G^2) \Biggr\} \; , \qquad \\
\Psi(t,r) &\!\!\! = \!\!\!& -\frac{GM}{ar} \Biggl\{1 + \frac{103 G}{15 
\pi a^2 r^2} - \frac{8 G H^2}{\pi} \ln^3(a) + O(G^2) \Biggr\} . \qquad
\end{eqnarray}
Even though the loop-counting parameter $G H^2 \ltwid 10^{-10}$ is very 
small, the steady growth of $\ln[a(t)]$ must eventually cause perturbation
theory to break down if inflation persists for a large number of e-foldings.
Understanding what happens after that point requires some sort of 
nonperturbative resummation technique. And exploring the fascinating 
question of what effects might persist to the current epoch requires a 
technique that is not specialized to de Sitter but can be applied to
any geometry (\ref{geometry}) which has experienced a phase of primordial
inflation.

Starobinsky's stochastic formalism \cite{Starobinsky:1986fx,
Starobinsky:1994bd} can be proven to capture the leading logarithms of 
scalar potential models to all orders \cite{Tsamis:2005hd}, but the 
derivative interactions of quantum gravity invalidate the proof, and 
direct calculation reveals that the technique fails even at one loop
\cite{Miao:2008sp}. Nonlinear sigma models such as,
\begin{equation}
\mathcal{L} = -\frac12 \Bigl( 1 + \frac12 \lambda \Phi\Bigr)^2 
\partial_{\mu} \Phi \partial_{\nu} \Phi \; , \label{Single} 
\end{equation}
possess similar derivative interactions, and induce the same sorts of 
large logarithms on de Sitter, but without the complicating features of 
tensor indices and gauge fixing. Much work has been done on such models 
in order to better understand quantum gravity \cite{Tsamis:2005hd,
Kitamoto:2010et,Kitamoto:2011yx,Kitamoto:2018dek,Miao:2021gic,
Woodard:2023rqo}. It has recently been shown that the large logarithms 
of nonlinear sigma models can be resummed by combining a variant of 
Starobinksy's technique with a variant of the renormalization group
\cite{Miao:2021gic}. Applying the renormalization group does not concern 
us here but facilitating Starobinsky's formalism motivates the current 
study. 

One applies Starobinsky's technique to (\ref{Single}) by first writing 
the exact field equation,
\begin{equation}
\frac{\delta S[\Phi]}{\delta \Phi} = \Bigl(1 + \frac12 \lambda \Phi\Bigr)
\partial_{\mu} \Bigl[ \Bigl(1 + \frac12 \lambda \Phi\Bigr) \sqrt{-g} \,
g^{\mu\nu} \partial_{\nu} \Phi\Bigr] = 0 \; . \label{EOM}
\end{equation}
Now note that a constant scalar background represents a field strength
renormalization of the free propagator,
\begin{equation}
\Phi(x) = \Phi_0 \qquad \Longrightarrow \qquad \Bigl\langle \Omega
\Bigl\vert \Phi(x) \Phi(x') \Bigr\vert \Omega \Bigr\rangle =
\frac{i\Delta(x;x')}{(1 + \frac12 \lambda \Phi_0)^2} \; . \label{fstrength}
\end{equation}
We can define an effective force due to undifferentiated scalars by 
integrating out the differentiated fields from the interaction part of 
(\ref{EOM}) using (\ref{fstrength}),
\begin{equation}
-V'_{\rm eff}(\Phi_0) \sqrt{-g} = \Bigl(1 + \frac{\lambda}{2} \Phi_0\Bigr)
\partial_{\mu} \Bigl[ \frac{\lambda}{4} \sqrt{-g} \, g^{\mu\nu} 
\partial_{\nu} \langle \Omega \vert \Phi^2 \vert \Omega \rangle \Bigr]
= -\frac{\frac{\lambda}{4} \frac{d}{dt} (a^{D-1} \dot{A})}{1 + 
\frac{\lambda}{2} \Phi_0} \; , \label{eforce}
\end{equation}
where $A(t)$ is the first of the three correlators in (\ref{threeDs}).
Integrating (\ref{eforce}) gives the effective potential,
\begin{equation}
V_{\rm eff}(\Phi) = \ln\Bigl\vert 1 \!+\! \frac{\lambda}{2} \Phi\Bigr\vert
\!\times\! \frac1{2 a^{D-1}} \frac{d}{dt} \Bigl( a^{D-1} \dot{A} \Bigr)
= \ln\Bigl\vert 1 \!+\! \frac{\lambda}{2} \Phi\Bigr\vert \!\times\! -
\frac12 \square A \; , \label{Veff}
\end{equation}
where $\square$ is the covariant scalar d'Alembertian. 

At this stage the system has been reduced to a scalar potential model, for 
which Starobinsky's stochastic formalism can be applied to capture the 
leading logarithms. When this is done on de Sitter, the evolution of the 
scalar background consists, at leading logarithm order, of a ``classical'' 
part from the field rolling down its (infinite) potential well, plus the 
contribution from stochastic ``jitter'' which accelerates the roll-down 
\cite{Miao:2021gic},
\begin{equation}
\Bigl\langle \Omega \Bigl\vert \Phi(t,\vec{x}) \Bigr\vert \Omega 
\Bigr\rangle_{\rm leading} \!\!\!\!\! = \frac{2}{\lambda} \Biggl\{ \Bigl[1 
\!-\! \frac{\lambda^2 H^2}{8 \pi^2} \ln(a) \Bigr]^{\frac14} - 1 \Biggr\} - 
\frac{3 \lambda^3 H^4}{2^8 \pi^4} \ln^2(a) + O(\lambda^5) \; . \label{VEVPhi} 
\end{equation}
Expression (\ref{VEVPhi}) has been explicitly verified at 1-loop and 2-loop 
orders \cite{Miao:2021gic}. If we knew the function $A(t)$ for a general
cosmology (\ref{geometry}) it would be possible to extend this result to
the current epoch, and even incorporate the back-reaction from the 
stress-energy of the field $\Phi$. Facilitating such analyses is the point
of this paper.

It is instructive to review how the correlators (\ref{threeDs}) look on de 
Sitter background with $\epsilon(t) = 0$. It was early realized that the 
expectation value of $\varphi^2$ on de Sitter background experiences secular 
growth \cite{Vilenkin:1982wt,Linde:1982uu,Starobinsky:1982ee},
\begin{equation}
\Bigl\langle \Omega \Bigl\vert \Phi^2(x) \Bigr\vert \Omega \Bigr
\rangle = \Bigl({\rm UV\ constant}\Bigr) + \frac{H^2}{4\pi^2} 
\ln\Bigl[a(t)\Bigr] \; . \label{olddS}
\end{equation}
In dimensional regularization the propagator is \cite{Onemli:2002hr,
Onemli:2004mb},
\begin{equation}
i\Delta(x;x') = F\Bigl(y(x;x')\Bigr) + k\Bigl[\ln(a a') - \pi 
{\rm cot}\Bigl(\frac{\pi D}{2} \Bigr)\Bigr] \;\; , \;\; k \equiv 
\frac{H^{D-2}}{(4\pi)^{\frac{D}2}} \frac{\Gamma(D\!-\!1)}{\Gamma(\frac{D}2)} 
\; , \label{dSprop}
\end{equation}
where $D$ is the dimension of spacetime and $y(x;x')$ is the de Sitter 
length function,
\begin{equation}
y(x;x') \equiv a a' \Bigl[H^2 \Vert \vec{x} - \vec{x}'\Vert^2
- \Bigl( \vert e^{-H t'} \!-\! e^{-H t} \vert \!-\! i \varepsilon\Bigr)^2 
\Bigr] \; , \label{ydef}
\end{equation}
and the function $F(y)$ is,
\begin{eqnarray}
\lefteqn{F(y) = \frac{H^{D-2}}{(4\pi)^{\frac{D}2}} \Biggl\{ 
\frac{\Gamma(\frac{D}2)}{\frac{D}2 \!-\! 1} \Bigl( \frac{4}{y}\Bigr)^{\frac{D}2 - 1}
+ \frac{\Gamma(\frac{D}2 \!+\! 1)}{\frac{D}{2} \!-\! 2} \Bigl(\frac{4}{y}
\Bigr)^{\frac{D}2 - 2} } \nonumber \\
& & \hspace{1.2cm} - \sum_{n=1}^{\infty} \Biggl[ \frac{\Gamma(n \!+\! \frac{D}2 
\!+\! 1)}{(n \!-\! \frac{D}2 \!+\! 2) \Gamma(n \!+\! 2)} \Bigl( \frac{y}{4}
\Bigr)^{n-\frac{D}2 + 2} - \frac{\Gamma(n \!+\! D \!-\! 1)}{n \Gamma(n \!+\!
\frac{D}2)} \Bigl( \frac{y}{4}\Bigr)^n \Biggr] \Biggr\} . \label{Adef} \qquad
\end{eqnarray}
On de Sitter background the fully dimensionally regulated coincidence limits
whose generalizations we seek are,
\begin{eqnarray}
i\Delta(x;x') \Bigl\vert_{x'=x} &\!\!\! = \!\!\!& -k \pi {\rm cot}\Bigl(
\frac{D \pi}{2}\Bigr) + 2 k \ln(a) \; , \qquad \label{dSpropfull} \\
\partial_{\mu} i\Delta(x;x') \Bigl\vert_{x' = x} &\!\!\! = \!\!\!& +k H 
u_{\mu} \longrightarrow +\frac{H^3}{8\pi^2} u_{\mu} \; ,
\qquad \label{dSprop'} \\
\partial_{\mu} \partial'_{\nu}i\Delta(x;x') \Bigl\vert_{x' = x} &\!\!\! = \!\!\!&
-\Bigl( \frac{D\!-\!1}{D}\Bigr) k H^2 g_{\mu\nu} \longrightarrow - 
\frac{3 H^4}{32 \pi^2} g_{\mu\nu} \; , \qquad \label{dSprop''}
\end{eqnarray}
where $u_{\mu}$ is a normalized, timelike 4-velocity --- $u_{\mu} = 
\delta^0_{~\mu}$ in the co-moving coordinates of (\ref{geometry}). The
purpose of this paper is to infer how expressions (\ref{dSpropfull}-\ref{dSprop''})
change when the first slow roll parameter is nonzero and the Hubble parameter
is time dependent.

Dolgov and Pellicia used the free scalar field equation to derive an
important relation between $A(t)$ and $C_{\mu\nu}(t)$ for a general
background geometry \cite{Dolgov:2005se},
\begin{equation}
\partial_{\mu} \Bigl[ \sqrt{-g} \, g^{\mu\nu} \partial_{\nu} A(t)\Bigr] =
-\frac{d}{dt} \Bigl( a^{D-1}(t) \dot{A}(t)\Bigr) = 2 \sqrt{-g} \, g^{\mu\nu} 
C_{\mu\nu}(t) \; . \label{Sasha1}
\end{equation}
Note that this relation applies to the {\it dimensionally regulated and 
unrenormalized} propagator. Of course we also have,
\begin{equation}
\partial_{\mu} A(t) = u_{\mu} \!\times\! \dot{A}(t) = 2 B_{\mu}(x) \; . 
\label{Sasha2}
\end{equation}
Homogeneity and isotropy expresses $C_{\mu\nu}(t)$ in terms of two 
functions,
\begin{equation}
C_{\mu\nu}(t) = u_{\mu} u_{\nu} \!\times\! C_0(t) + \overline{g}_{\mu\nu}
\!\times\! \overline{C}(t) \qquad , \qquad \overline{g}_{\mu\nu} \equiv
g_{\mu\nu} + u_{\mu} u_{\nu} \; . \label{Sasha3}
\end{equation}
Finally, conservation of the massless, minimally coupled scalar stress 
tensor relates $C_0(t)$ and $\overline{C}(t)$,
\begin{equation}
\frac{\dot{C}_0}{D \!-\! 1} + \dot{\overline{C}} + 2 H \Bigl[C_0 + 
\overline{C}\Bigr] = 0 \; . \label{Sasha4}
\end{equation}
So $A$ determines $B_{\mu}$ and $C_A \equiv -C_0 + (D-1) \overline{C}$,
and conservation (\ref{Sasha4}) gives the other linear combination $C_B 
\equiv C_0 + \overline{C}$.

The goal of this paper is to determine a good analytic approximation 
for $A(t)$ for a cosmological geometry (\ref{geometry}) which has 
experienced a phase of primordial inflation. We will not renormalize
but instead employ dimensional regularization to derive {\it primitive}
results for the three correlators (\ref{threeDs}), which are exact for
the divergent parts and good approximations for the finite parts. This
leaves readers free to apply whatever renormalization conditions they
wish. In section 2 we express $A(t)$ as a spatial Fourier mode sum of 
an amplitude $\mathcal{A}(t,k)$, and we define a plausible geometry 
which incorporates both primordial inflation and a subsequent 
$\Lambda$CDM expansion history. Section 3 develops analytic 
approximations for $\mathcal{A}(t,k)$ before first horizon crossing, 
between first and second crossings, and after second crossing. Although 
our approximations are valid for {\it any} geometry (\ref{geometry}) 
which has undergone primordial inflation, we test them against numerical 
evolution in the plausible geometry. In section 4 we evaluate the 
spatial Fourier mode sum to obtain explicit, analytic results for 
$A(t)$, both during primordial inflation and afterwards. The quantities 
$B_{\mu}(t)$, $C_0(t)$ and $\overline{C}(t)$ are derived in section 5. 
Our conclusions comprise section 6.

\section{The Amplitude $\mathcal{A}(t,k)$}

The purpose of this section is to give an exact expression for $A(t)$ 
as a dimensionally regulated, spatial Fourier mode sum of an amplitude
$\mathcal{A}(t,k)$ and to derive equations governing this amplitude.
We also devise a geometry which interpolates between an early phase 
of primordial inflation and the current $\Lambda$CDM expansion history. 
In future sections we compare analytic approximations for $\mathcal{A}(t,k)$
with numerical evolution in this geometry. The section closes by giving a 
dimensionless formulation which is appropriate for numerical evolution.

\subsection{Preliminaries}

The $\Phi(x)$ propagator can be expressed as the inverse Fourier transform 
(regulated in $D$ spacetime dimensions) of plane wave mode functions $u(t,k)$,
\begin{eqnarray}
\lefteqn{i\Delta(x;x') = \int \!\! \frac{d^{D-1}k}{(2\pi)^{D-1}} \Biggl\{ 
\theta(t \!-\!t') u(t,k) u^*(t',k) e^{i \vec{k} \cdot (\vec{x} - \vec{x}')} }
\nonumber \\
& & \hspace{5.5cm} + \theta(t' \!-\!t) u^*(t,k) u(t',k) e^{-i \vec{k} \cdot (
\vec{x} - \vec{x}')} \Biggr\} . \qquad \label{Ftrans}
\end{eqnarray}
The mode functions obey the equations,
\begin{equation}
\ddot{u} + (D \!-\! 1)  H \dot{u} + \frac{k^2}{a^2} u = 0 \qquad , \qquad 
u \dot{u}^* - \dot{u} u^* = \frac{i}{a^{D-1}} \; . \label{ueqns}
\end{equation}
Although the mode equation cannot be solved for general $a(t)$, the ultraviolet
solution is,
\begin{equation}
k \gg H(t) a(t) \qquad \Longrightarrow \qquad u(t,k) \longrightarrow 
\frac1{\sqrt{2 k a^{D-2}(t)}} \, \exp\Bigl[ i k \!\! \int_{t_i}^{t} \!\! 
\frac{dt'}{a(t')} \Bigr] \; . \label{WKBform}
\end{equation}
We will use this to infer initial conditions.

Equation (\ref{Ftrans}) implies that the function $A(t)$ can be expressed
as a spatial Fourier mode sum of the amplitude $\mathcal{A}(t,k) \equiv 
u(t,k) u^*(t,k)$,
\begin{equation}
A(t) = \int \!\! \frac{d^{D-1}k}{(2\pi)^{D-1}} \, \mathcal{A}(t,k) \; . 
\label{Asum}
\end{equation}
The other coincidence limits have similar expressions,
\begin{eqnarray}
B_{\mu}(t) &\!\!\! = \!\!\!& \delta^0_{~\mu} \!\! \int \!\! 
\frac{d^{D-1}k}{(2\pi)^{D-1}} \, \dot{\mathcal{A}}(t,k) \; , \qquad
\label{Bsum} \\
C_0(t) &\!\!\! = \!\!\!& \int \!\! 
\frac{d^{D-1}k}{(2\pi)^{D-1}} \Bigl[\frac12 \ddot{\mathcal{A}}(t,k) +
\frac{(D\!-\!1)}{2} H \dot{\mathcal{A}}(t,k) + \frac{k^2}{a^2(t)} 
\mathcal{A}(t,k)\Bigr] \; , \qquad \label{C0sum} \\
\overline{C}(t) &\!\!\! = \!\!\!& \int \!\! 
\frac{d^{D-1}k}{(2\pi)^{D-1}} \frac{k^2}{a^2(t)} \mathcal{A}(t,k) \; .
\qquad \label{Cbarsum}
\end{eqnarray}
Note that while $A(t)$ and $B_{\mu}(t)$ are only quadratically divergent,
$C_0(t)$ and $\overline{C}(t)$ diverge quartically.

After some manipulations (for all the details, see \cite{Romania:2012tb,
Brooker:2015iya}) the mode equation and the Wronskian (\ref{ueqns}) can be 
combined to give a nonlinear, 2nd order equation for $\mathcal{A}(t,k)$,
\begin{equation}
\ddot{\mathcal{A}} - \frac{\dot{\mathcal{A}}^2}{2 \mathcal{A}} + (D\!-\!1) H 
\dot{\mathcal{A}} + \frac{2 k^2}{a^2} \mathcal{A} - \frac1{2 a^{2D-2} \mathcal{A}}
= 0 \; . \label{Aeqnnonlinear}
\end{equation}
And a 3rd order, linear equation can also be derived,
\begin{equation}
\dddot{\mathcal{A}} + 3 (D\!-\!1) H \ddot{A} + (D\!-\!1) [2 (D\!-\!1) H^2 
+ \dot{H}] \dot{\mathcal{A}} + \frac{4 k^2}{a^2} \dot{\mathcal{A}} + 
\frac{4 (D\!-\!2) H k^2}{a^2} \mathcal{A} = 0 \; . \label{Aeqnlinear}
\end{equation}
The WKB form (\ref{WKBform}) implies initial conditions,
\begin{equation}
\mathcal{A}(t_i,k) = \frac1{2 k a_i^{D-2}} \;\; , \;\; \dot{\mathcal{A}}(t_i,k)
= -\frac{(D\!-\!2) H_i}{2 k a_i^{D-2}} \;\; , \;\; \ddot{\mathcal{A}}(t_i,k)
= \frac{(D\!-\!2)^2 H_i^2}{2 k a_i^{D-2}} \; . \label{Ainitial}
\end{equation}

\subsection{A Plausible Expansion History}

We assume that the massless, minimally coupled scalar $\Phi(x)$ is 
initially a spectator to primordial inflation driven by a minimally coupled 
scalar inflaton $\varphi(t)$ with potential $V(\varphi)$. The nontrivial 
Einstein equations are,
\begin{eqnarray}
(D\!-\!1) H^2 & = & 8\pi G \Bigl[ \frac12 \dot{\varphi}^2 + V(\varphi) 
\Bigr] \; , \label{Infl1} \\
- [(D\!-\!1) \!-\! 2 \epsilon] H^2 & = & 8 \pi G \Bigl[ \frac12 \dot{\varphi}^2 
- V(\varphi) \Bigr] \; . \label{Infl2}
\end{eqnarray}
The inflaton itself evolves according to the equation,
\begin{equation}
\ddot{\varphi} + (D\!-\!1) H \dot{\varphi} + V'(\varphi) = 0 \; . \label{Infl3}
\end{equation}
Combining (\ref{Infl1}) and (\ref{Infl2}) gives the first slow roll parameter,
\begin{equation}
\epsilon_{\varphi}(t) = \frac{ (\frac{D-1}{2}) \dot{\varphi}^2}{\frac12 
\dot{\varphi}^2 + V(\varphi)} \; . \label{infleps}
\end{equation}

Of course numerical evolution requires a specific potential. For this we have 
chosen the simple quadratic model,
\begin{equation}
V(\varphi) = \frac{c^2 \varphi^2}{16 \pi G} \; . \label{potential}
\end{equation}
Taking the initial value of the scalar to be $\varphi_i = 15/\sqrt{8\pi G}$ results 
in about 56.8 e-foldings of inflation. Although this model is inconsistent with the 
current bound on the tensor-to-scalar ratio,\footnote{The flatter potentials of more
realistic models cause $\epsilon$ to be smaller, and more nearly constant, both of
which serve to make our analytic approximations even more accurate.} choosing 
$c = 7.1 \times 10^{-6}$ results in the correct scalar amplitude and spectral index 
\cite{Planck:2018vyg}. 

Inflation ends when $\epsilon(t_e) = 1$, after which $\epsilon(t)$ oscillates between
$\epsilon = 0$ and $\epsilon = 3$, with constant amplitude and increasing frequency.
A realistic model would couple the inflaton to normal matter, which would be heated
by the inflaton's kinetic energy to produce a radiation-dominated universe. At late
times we want to reach the $\Lambda$CDM model,
\begin{equation}
H^2 = H_0^2 \Bigl[ \Omega_r (1 \!+\! z)^4 + \Omega_m (1 \!+\! z)^3 + \Omega_{\Lambda}
\Bigr] \qquad , \qquad 1 + z \equiv \frac{a_0}{a(t)} \; , \label{LCDM}
\end{equation}
where the subscript $0$ represents the current time and the various parameters are
\cite{Planck:2018vyg},
\begin{equation}
H_0 \simeq 67.3 \, \frac{\rm km/s}{\rm Mpc} \quad , \quad \Omega_r \simeq 
\frac{\Omega_m}{3390} \quad , \quad \Omega_m \simeq 0.315 \quad , \quad \Omega_{\Lambda}
\simeq 0.685 \; . \label{LCDMnumbers}
\end{equation}
\begin{figure}[H]
\centering
\includegraphics[width=6.5cm]{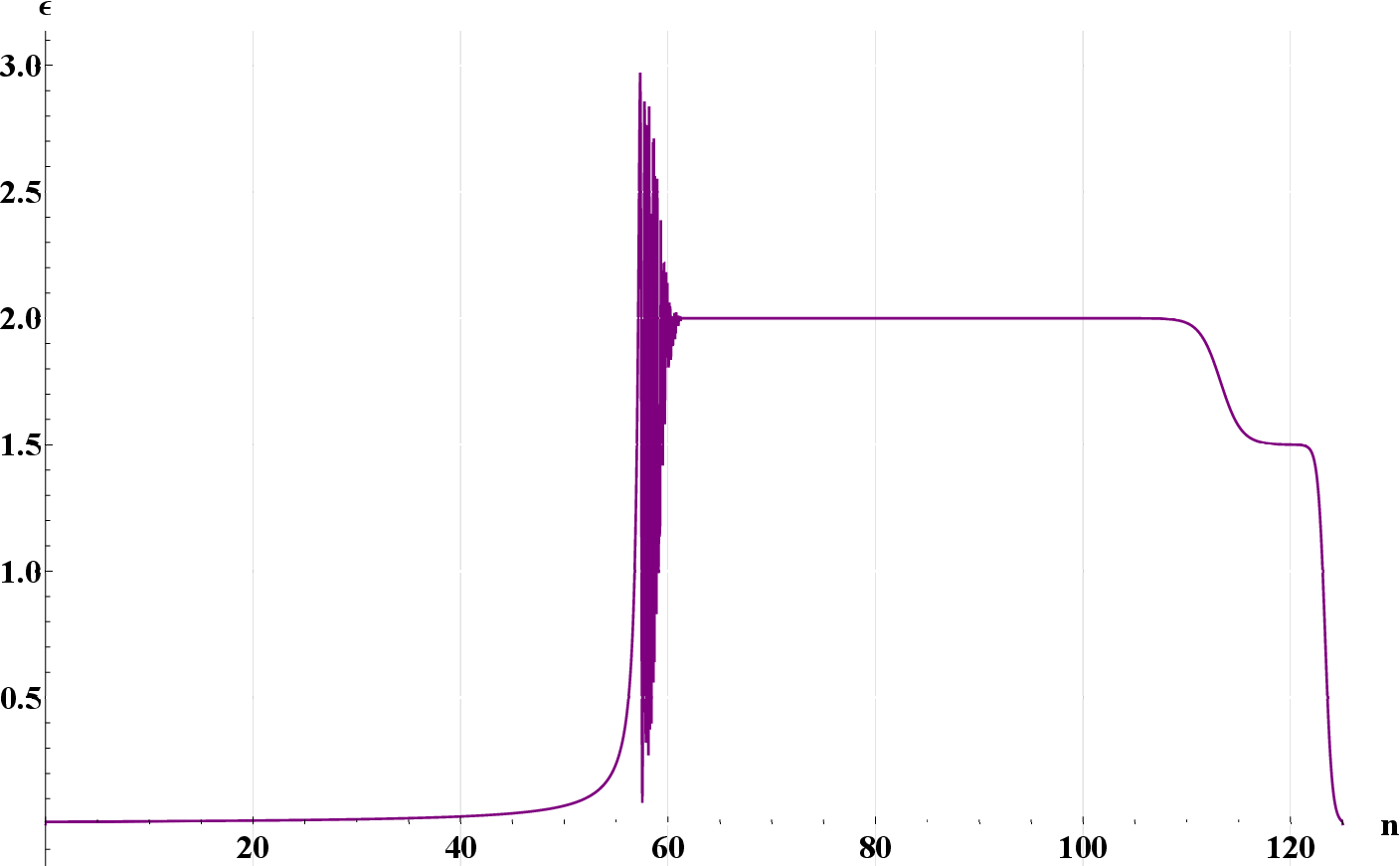}
\hskip .5cm
\includegraphics[width=6.5cm]{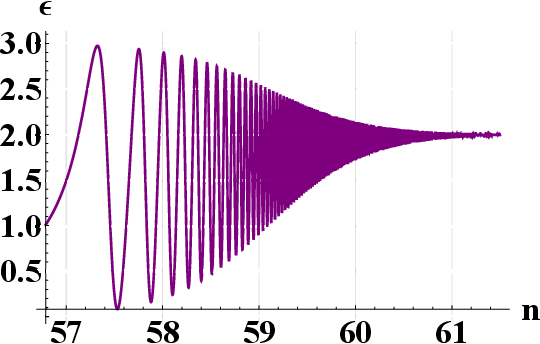}
\caption{\footnotesize The first plot shows first slow roll parameter $\epsilon$
over the full range of cosmic history. The second plot the interpolation 
(\ref{fulleps}) during reheating.}
\label{epsilonfull}
\end{figure}
\noindent The first slow roll parameter of the $\Lambda$CDM model is,
\begin{equation}
\epsilon_{\Lambda}(t) = \frac{2 \Omega_r (1 \!+\! z)^4 + \frac32 \Omega_m 
(1 \!+\! z)^3}{\Omega_r (1 \!+\! z)^4 + \Omega_m (1 \!+\! z)^3 + \Omega_{\Lambda}} 
\qquad , \qquad 1 \!+\! z = \frac{a_0}{a(t)} \; . \label{LCDMeps}
\end{equation}
Rather than devise a model of reheating, we have merely made an {\it ad hoc} interpolation 
of the first slow roll parameter between the scalar-driven result (\ref{infleps}) and the
$\Lambda$CDM result (\ref{LCDMeps}),
\begin{equation}
\epsilon(t) = \frac12 \Bigl[1 - \tanh(n \!-\! n_{\rm eq})\Bigr] \!\times\! 
\epsilon_{\varphi}(t) + \frac12 \Bigl[1 + \tanh(n \!-\! n_{\rm eq} )\Bigr] \!\times\! 
\epsilon_{\Lambda}(t) \; , \label{fulleps}
\end{equation}
where $\epsilon_{\varphi}(t)$ and $\epsilon_{\Lambda}(t) \simeq 2$ were defined in 
expressions (\ref{infleps}) and (\ref{LCDMeps}), respectively, $n$ is the number 
of e-foldings since the start of inflation at $t = t_i$, $n \equiv \ln[a(t)/a(t_i)]$.
Figure~\ref{epsilonfull} shows the resulting geometry.

The e-folding $n_{\rm eq} = 59.1$ at which scalar-driven inflation and the 
$\Lambda$CDM model are equally represented was chosen for numerical convenience.
Any value more than a few e-foldings after the end of inflation at $n_e \simeq 56.8$
could be used instead. The current time $t_0$, corresponding to $n_0 \simeq 123.9$, 
is found by integrating expression (\ref{fulleps}) --- using $\epsilon_{\Lambda} = 2$ 
--- from the end of inflation to a time $t_{\rm rad}$ a few e-foldings past $n = 
n_{\rm eq}$. By that point the $\Lambda$CDM geometry dominates, and we can assume its 
radiation-dominated form to write,
\begin{equation}
H(t_{\rm rad}) = \frac{H(t_e)}{1 + H(t_e) \int_{t_e}^{t_{\rm rad}} \!\! dt' \,
\epsilon(t') } \equiv H_0 \sqrt{\Omega_{r}} e^{2 (n_{\rm rad} - n_0)} \; . 
\label{n0def}
\end{equation} 

We close by stressing that this geometry has been introduced merely to provide an 
explicit framework for comparing the exact numerical evolution of $\mathcal{A}(t,k)$ 
with the analytic approximations we shall develop in section 3. These analytic 
approximations have nothing to do with specific properties of our plausible geometry.
In particular, they do not depend on inflation being supported by a scalar potential,
much less the quadratic potential, nor do they depend on the initial value of this 
scalar, the constant $c$, or the interpolation (\ref{fulleps}) between scalar-driven 
inflation and the $\Lambda$CDM cosmology.

\subsection{Dimensionless Variables}

The variable $n \equiv \ln[a(t)/a_i]$ is preferable to $t$, both because $n$ is 
dimensionless and because it is less sensitive to dramatic changes which take 
place in the time scale of events as inflation progresses. Derivatives obey,
\begin{equation}
\frac{d}{dt} = H \frac{d}{dn} \; , \; \frac{d^2}{d t^2} = H^2 \Bigl[ 
\frac{d^2}{d n^2} \!-\! \epsilon \frac{d}{d n} \Bigr] \; , \; \frac{d^3}{dt^3}
= H^3 \Bigl[ \frac{d^3}{d n^3} \!-\! 3 \epsilon \frac{d^2}{d n^2} \!+\! (2 \epsilon^2
\!-\! \epsilon') \frac{d}{d n}\Bigr] . \label{timeton}
\end{equation}
Just as dots denote differentiation with respect to $t$ we use primes to stand for
differentiation with respect to $n$, except that a primed potential still represents
its derivative with respect to the scalar. It is convenient to factor the dimensions 
out of the inflaton, the Hubble parameter and the inflaton potential,
\begin{equation}
\psi(n) \equiv \sqrt{8 \pi G} \, \varphi(t) \quad , \quad \chi(n) \equiv \sqrt{8\pi G} 
\, H(t) \quad , \quad U(\psi) \equiv (8\pi G)^2 V(\varphi) \; . \label{dimless1}
\end{equation}
Of course the first slow roll parameter $\epsilon = -\chi'/\chi$ is already 
dimensionless. Using these variables, and expressions (\ref{Infl1}-\ref{Infl2}), to
solve for the geometrical quantities in terms of the scalar,
\begin{equation}
\chi^2 = \frac{U}{3 \!-\! \frac12 {\psi'}^2} \qquad , \qquad \epsilon = \frac12 
{\psi'}^2 \; . \label{Infl4}
\end{equation}
We can also re-express the scalar evolution equation (\ref{Infl3}) as,
\begin{equation}
\psi'' + (3 \!-\! \epsilon) \psi' + \frac{U'(\psi)}{\chi^2} = 0 \qquad \Longrightarrow 
\qquad \psi'' + \Bigl(3 - \frac12 {\psi'}^2\Bigr) \Bigl[ \psi' + \frac{U'}{U} \Bigr]
= 0 \; . \label{Infl5}
\end{equation}
For the quadratic potential $U(\psi) = \frac12 c^2 \psi^2$ we have chosen, the
slow roll approximation gives,
\begin{equation}
\psi(n) \simeq \sqrt{\psi_i^2 \!-\! 4n} \quad , \quad \chi(n) \simeq \frac{c}{\sqrt{6}} 
\sqrt{\psi_i^2 \!-\! 4n} \quad , \quad \epsilon \simeq \frac{2}{\psi_i^2 \!-\! 4 n}
\; . \label{slowroll}
\end{equation}

We scale out the dimensions of the wave number and the amplitude,
\begin{equation}
\kappa \equiv \sqrt{8 \pi G} \, k \qquad , \qquad \alpha(n,\kappa) \equiv 
\frac{\mathcal{A}(t,k)}{\sqrt{8\pi G}} \; . \label{dimeless2}
\end{equation}
The 2nd order and 3rd order equations (\ref{Aeqnnonlinear}-\ref{Aeqnlinear}) become, 
\begin{eqnarray}
0 &\!\!\! = \!\!\!& \alpha'' - \frac{{\alpha'}^2}{2 \alpha} + (D\!-\!1\!-\!\epsilon) 
\alpha' + \frac{2 \kappa^2 \alpha}{e^{2n} \chi^2} - \frac{e^{-2 (D-1) n}}{2 \chi^2 
\alpha} \; , \qquad \label{alphaeqnnonlinear} \\
0 &\!\!\! = \!\!\!& \alpha''' + 3 (D \!-\! 1 \!-\! \epsilon) \alpha'' + 
\Bigl[ 2 (D\!-\!1 \!-\! \epsilon)^2 \!-\! \epsilon'\Bigr] \alpha' \nonumber \\
& & \hspace{7.5cm} + \frac{4 \kappa^2 \alpha'}{e^{2n} \chi^2} + \frac{4 (D\!-\!2) 
\kappa^2 \alpha}{e^{2n} \chi^2} \; . \qquad \label{alphaeqnlinear}
\end{eqnarray}
The WKB form (\ref{WKBform}) implies initial conditions,
\begin{equation}
\alpha(0,\kappa) = \frac1{2\kappa} \quad , \quad \alpha'(0,\kappa) = -
\frac{(D\!-\!2)}{2 \kappa} \quad , \quad \alpha''(0,\kappa) = 
\frac{(D\!-\!2)^2}{2 \kappa} \; . \label{initialalpha}
\end{equation}

\section{Approximating $\mathcal{A}(t,k)$}
 
The purpose of this section is to develop analytic approximations for the
amplitude $\mathcal{A}(t,k)$ according to where the wave number $k$ lies
with respect to $H(t) a(t)$. When $k = H a$ the mode is said to experience 
{\it horizon crossing}. The section begins with a discussion of horizon
crossing. We then give successive analytic approximations for ultraviolet 
modes which have never experienced horizon crossing, for modes which have 
experienced one crossing and for modes which have twice experienced 
crossing.

\subsection{Horizon Crossing}

Modes are said to be {\it sub-horizon} if $k > H(t) a(t)$ and {\it 
super-horizon} if $k < H(t) a(t)$. We do not assume that the mode sums 
(\ref{Asum}-\ref{Cbarsum}) run all the way down to $k = 0$, but rather that 
the far infrared portion is cut off for modes $k < k_i = H(t_i) a(t_i)$ 
which were super-horizon at the beginning of inflation. (Justified by
pre-inflationary modes being less highly excited \cite{Vilenkin:1983xp},
or else by the spatial manifold being compact \cite{Tsamis:1993ub}.)
This means that all modes are initially sub-horizon and may, or may not, 
experience horizon crossing during the course of primordial inflation. 
Some of those modes which have experienced first horizon crossing may 
experience second crossing. 

\begin{figure}[H]
\centering
\includegraphics[width=8cm]{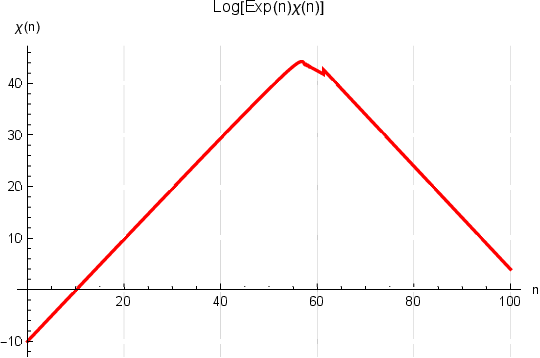}
\caption{\footnotesize Behavior of $\ln[\chi(n) e^{n}]$ over the course of
cosmic history.} \label{horizon}
\end{figure}

Figure~\ref{horizon} shows the evolution of the logarithm of $\chi(n) e^{n} 
= \sqrt{8\pi G} \times H(t) a(t)$ over the course of the plausible expansion
history described in section 2.2.
\noindent We define the first horizon crossing time $t_k$ as the time when a 
mode with wave number $k$ intersects the left hand, rising portion of the curve, 
$k = a(t_k) H(t_k)$. During primordial inflation the product $H(t) a(t)$ grows 
from $k_i$ and distinguishes modes which have experienced first horizon crossing
from those which are still ultraviolet,  
\begin{equation}
{\rm During\ Inflation} \qquad \Longrightarrow \qquad k_i < H(t) a(t) < \infty 
\; . \label{ksduring}
\end{equation}
If $t_e$ represents the end of primordial inflation then, except for a brief
oscillatory period during reheating, $H(t) a(t)$ falls off from $k_e \equiv 
H(t_e) a(t_e)$ and we can distinguish a 3rd class of modes --- between the 
infrared and the ultraviolet --- which have experienced 2nd horizon crossing,
\begin{equation}
{\rm After\ Inflation} \qquad \Longrightarrow \qquad k_i < H(t) a(t) < k_e
< \infty \; . \label{ksafter}
\end{equation} 

In performing the mode sum (next section) it will be desirable to change 
variables from wave number $k$ to the first horizon crossing time $t_k$,
\begin{equation}
k = a(t_k) H(t_k) \qquad \Longrightarrow \qquad \frac{dk}{k} = [1 \!-\! 
\epsilon(t_k)] H(t_k) dt_k \; . \label{ktotk}
\end{equation}
It is also useful to have an expression for the time $t_2(t_k)$ that a mode
with first horizon crossing time experiences second horizon crossing,
\begin{equation}
k \equiv a(t_k) H(t_k) \equiv a\Bigl(t_2(t_k)\Bigr) H\Bigl(t_2(t_k)\Bigr)
\; . \label{t2def}
\end{equation} 
Figure~\ref{horizon} shows that this is a well-defined function, except 
for the late phase of cosmic acceleration, and the small oscillatory range 
during reheating. The inverse of $t_2(t_1)$ is $t_1(t_2)$,
\begin{equation}
t_1\Bigl(t_2(t)\Bigr) \equiv t \qquad \Longleftrightarrow \qquad 
t_1\Bigl(t_2(t)\Bigr) = t \; . \label{t1def}
\end{equation}

\subsection{Before 1st Horizon Crossing}

\begin{figure}[H]
\centering
\includegraphics[width=10cm]{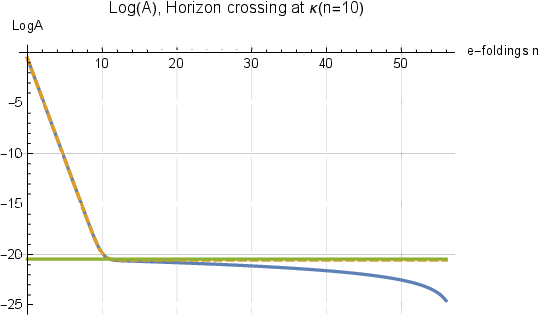}
\caption{\footnotesize Comparisons of the exact numerical evolution of 
$\ln[\alpha(n,\kappa)]$ (in dashed yellow) versus the asymptotic series of
expressions (\ref{gammazeta}) and (\ref{gammaexp}) (in solid blue) for a mode 
which experiences horizon crossing at $n = 10$. The horizontal solid green 
line shows the (logarithm of the dimensionless) freeze-in value 
(\ref{freezein}-\ref{Cdef}).} \label{hc10}
\end{figure}

The dimensionless amplitude $\alpha(n,\kappa) = \mathcal{A}(t,k)/\sqrt{8\pi G}$ 
is not known for general $\epsilon$ but it is easy to develop an ultraviolet 
expansion which applies for $\kappa \gg e^{n} \chi(n)$. To do this we make the 
definitions,
\begin{equation}
\alpha(n,\kappa) \equiv \frac{\gamma(n,\kappa)}{2 \kappa e^{(D-2)n}} \qquad , \qquad
\zeta(n,\kappa) \equiv \frac{\kappa}{e^n \chi(n)} \; , \label{gammazeta}
\end{equation}
and substitute into equation (\ref{alphaeqnnonlinear}) to obtain,
\begin{equation}
\gamma'' - \frac{{\gamma'}^2}{2 \gamma} + (1 \!-\! \epsilon) \gamma' - \frac12 
(D\!-\!2) (D \!-\! 2 \epsilon) \gamma + 2 \zeta^2 \Bigl( \gamma - \frac1{\gamma}\Bigr)
= 0 \; . \label{gammaeqn}
\end{equation}
Relation (\ref{gammaeqn}) gives an expansion for $\gamma(n,\kappa)$ in powers
of $\zeta^{-2}$, where the coefficients involve increasing numbers of derivatives
of $\epsilon(n)$,
\begin{eqnarray}
\lefteqn{\gamma = 1 + \frac{(D\!-\!2) (D\!-\!2\epsilon)}{8 \zeta^2}
+ \frac{3 (D\!-\!2) (D \!-\! 2 \epsilon) (D \!-\! 4 \!+\! 2 \epsilon) (D \!+\! 2
\!-\! 4 \epsilon)}{128 \zeta^4} + \dots } \nonumber \\
& & \hspace{5cm} + \frac{(D \!-\! 2) [(D \!+\! 5 \!-\! 7 \epsilon) \epsilon' \!+\!
\epsilon'']}{16 \zeta^4} + O(\zeta^{-6}) \; . \qquad \label{gammaexp}
\end{eqnarray}
Note that $\gamma = 1$ is an exact solution for $D = 2$ and/or for perfect
radiation-domination ($\epsilon = \frac{D}2$). Figure~\ref{hc10} compares
the exact numerical evolution with the asymptotic series (including the
$\zeta^{-2}$ contribution in (\ref{gammaexp}) but not the $\zeta^{-4}$ term)
for a mode which experiences horizon crossing at $n=10$. It is not even 
possible to discern any difference between the two until after horizon 
crossing.

\subsection{Between 1st and 2nd Horizon Crossing}

Because ultraviolet divergences are associated with $k \rightarrow \infty$, we 
can suspend dimensional regularization for modes which have experienced first 
horizon crossing. If the modes have not yet experienced second horizon crossing
then the last two terms of equation (\ref{Aeqnnonlinear}) are negligible and
we see that $\mathcal{A}(t,k)$ approaches a constant. This constant is
approximately \cite{Brooker:2015iya},
\begin{equation}
t > t_k \Longrightarrow \mathcal{A}(t,k) \simeq \frac{H^2(t_k) C(\epsilon_k)}{2 k^3}
\Biggl\{1 + O\Bigl( \frac{k^2}{a^2(t) H^2(t)} \Bigr) \Biggr\} \; , \label{freezein}
\end{equation}
where the function $C(\epsilon)$ is,
\begin{equation}
C(\epsilon) \equiv \frac1{\pi} \Gamma^2\Bigl(\frac12 + \frac1{1 \!-\! \epsilon}
\Bigr) \Bigl[2 (1 \!-\! \epsilon)\Bigr]^{\frac2{1-\epsilon}} \; . \label{Cdef} 
\end{equation}
Expressions (\ref{freezein}-\ref{Cdef}) give the freeze-in amplitude to all orders
in the slow roll approximation. 

The solid green line in Figure~\ref{hc10} gives the logarithm of the freeze-in 
amplitude (\ref{freezein}-\ref{Cdef}) for a mode which experiences horizon 
crossing at $n = 10$. It is difficult to detect any difference between it and the
numerical solution (in dashed yellow) after horizon crossing. We possess good 
analytical approximations for additional, nonlocal contributions to the freeze-in
amplitude, but these are very small unless the first slow roll parameter varies 
wildly near the time of first horizon crossing \cite{Brooker:2017kjd}.

\subsection{After 2nd Horizon Crossing}

\begin{figure}[H]
\centering
\includegraphics[width=6cm]{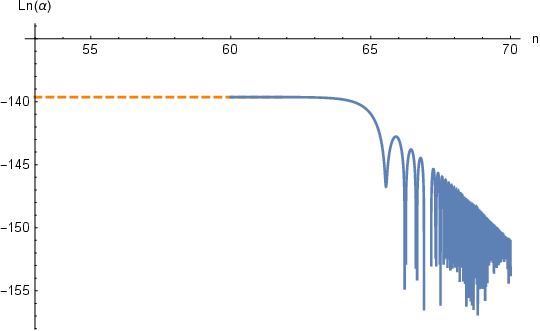}
\hskip 1cm
\includegraphics[width=6cm]{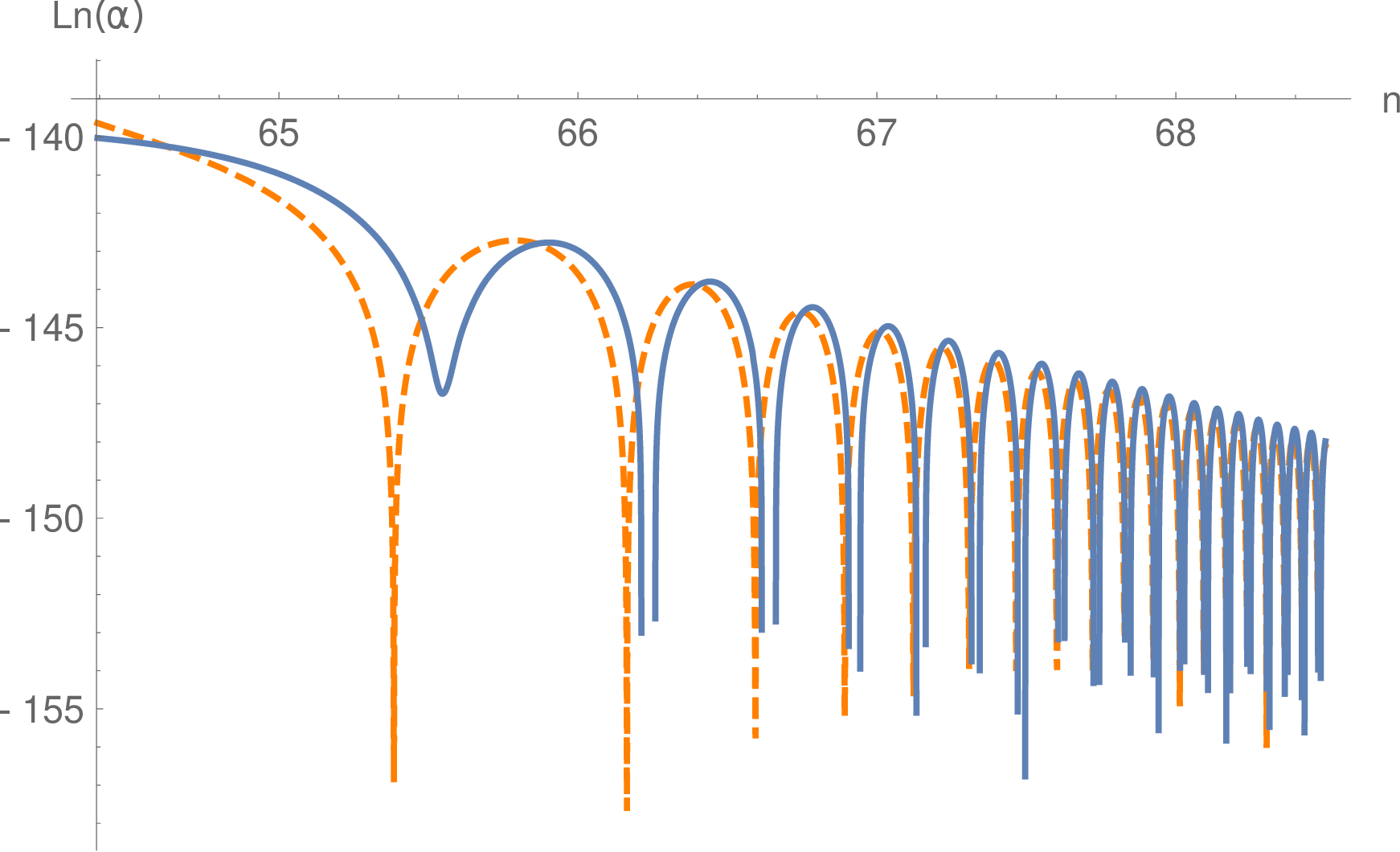}
\caption{\footnotesize The left hand plot shows the numerical solution (in
solid blue) of $\ln[\alpha(n,\kappa)]$ for a mode which experiences first 
horizon crossing at $n = 50$, and then experiences second horizon crossing 
at $n \simeq 64.5$. The yellow dashed line on the left gives the freeze-in
value (\ref{freezein}). The right hand plot compares the exact numerical
solution (in solid blue) of $\ln[\alpha(n,\kappa)]$ with the analytic 
approximation (\ref{post2}) (in dashed yellow dashed).}
\label{2ndcross}
\end{figure}

After 2nd horizon crossing the time dependence of the WKB form (\ref{WKBform}),
applies but starting with $\dot{\mathcal{A}}(t,k) = 0$ at $t = t_2(t_k)$ from the 
freeze-in amplitude (\ref{freezein}-\ref{Cdef}),
\begin{equation}
t > t_2(t_k) \Longrightarrow \mathcal{A}(t,k) \simeq \frac{H^2(t_k) C(\epsilon(t_k))}{
2 k^3} \!\times\! \Bigl[\frac{a(t_2(t_k))}{a(t)}\Bigr]^2 \! \cos^2\Biggl[ 
\int_{t_2(t_k)}^{t} \!\!\!\!\!\!\!\! dt' \frac{k}{a(t')} \Biggr] . \label{post2}
\end{equation}
Figure~\ref{2ndcross} compares the exact numerical evolution (on the left) with
the analytic approximation (on the right) for a mode which experiences first
horizon crossing at $n = 50$ and second horizon crossing at $n \simeq 64.5$. 
The agreement is quite good, except for the first oscillation which suffers 
from the usual inaccuracy when the WKB frequency vanishes. Because the rapid
oscillations of the cosine-squared average to $\frac12$ inside the mode sum,
the important part is the $1/a^2(t)$ damping.

\section{Results for $A(t)$}

The purpose of this section is to insert the approximate results 
(\ref{gammazeta}-\ref{gammaexp}), (\ref{freezein}-\ref{Cdef}) and
(\ref{post2}) for the amplitude $\mathcal{A}(t,k)$ into the mode sum
(\ref{Asum}) for $A(t)$. We begin with the case of primordial inflation,
during which the mode sum can be decomposed into before and after first
horizon crossing. After the end of inflation the modes which have
most recently experienced 1st crossing experience 2nd crossing, which
gives a third range of modes between the first two. We then consider
what the expectation value of the scalar propagator might be for a 
general metric. The section closes with a comment on why we do not 
employ the instantaneously constant $\epsilon$ approximation.

\subsection{During Primordial Inflation}

Because the amplitude $\mathcal{A}(t,k)$ depends on the wave vector 
$\vec{k}$ only through its magnitude, we can perform the angular 
integrations in the mode sum (\ref{Asum}),
\begin{equation}
A(t) = \frac{2}{\Gamma(\frac{D-1}{2}) (4\pi)^{\frac{D-1}{2}}} \!
\int_{k_i}^{\infty} \!\!\!\! dk \, k^{D-2} \mathcal{A}(t,k) \; .
\label{A1}
\end{equation}
Note that the lower limit of the mode sum has been cut off at the wave 
number $k_i = H(t_i) a(t_i)$ which is just crossing the horizon at the
start of inflation. During inflation we distinguish between ultraviolet 
modes ($H(t) a(t) < k < \infty$), which have not yet experienced first 
horizon crossing, and infrared ones ($k_i < k < H(t) a(t)$), which 
have already done so,
\begin{equation}
\int_{k_i}^{\infty} \!\!\!\! dk = \int_{H a}^{\infty} \!\!\!\!\!\!
dk + \int_{k_i}^{H a} \!\!\!\!\!\! dk \; . \label{intdkduring}
\end{equation}
Figure~\ref{hc10} shows that we only need the first two terms of the
ultraviolet expansion (\ref{gammazeta}-\ref{gammaexp}) to accurately
describe $\mathcal{A}(t,k)$ right up to the point of first horizon
crossing, and that the freeze-in amplitude (\ref{freezein}-\ref{Cdef}) 
is valid afterwards. Making the appropriate substitutions in the
mode sum gives,
\begin{eqnarray}
\lefteqn{A(t) \simeq \frac{2}{\Gamma(\frac{D-1}{2}) (4\pi)^{\frac{D-1}{2}}}
\int_{aH}^{\infty} \!\! \frac{ dk k^{D-2}}{2 k a^{D-2}} \Biggl\{1 + 
\frac{(D\!-\!2) (D\!-\!2\epsilon)}{8} \frac{a^2 H^2}{k^2}\Biggr\} }
\nonumber \\
& & \hspace{7cm} + \frac1{2\pi^2} \!\! \int_{a_i H_i}^{a H} \!\!\!\! dk \,
k^2 \frac{H^2(t_k) C(\epsilon_k)}{2 k^3} \; . \qquad \label{A2}
\end{eqnarray}
Note that we have suspended dimensional regularization on the infrared
portion of the mode sum.

Under the rules of dimensional regularization, any $D$-dependent power
of infinity (or zero) vanishes,
\begin{eqnarray}
\int_{aH}^{\infty} \!\!\!\! dk \, k^{D-3} & = & \frac{k^{D-2}}{D - 2} 
\Bigl\vert_{aH}^{\infty} = -\frac{(a H)^{D-2}}{D - 2} \longrightarrow 
-\frac12 (a H)^2 \; , \\
\int_{aH}^{\infty} \!\!\!\! dk \, k^{D-5} & = & \frac{k^{D-4}}{D - 4} 
\Bigl\vert_{aH}^{\infty} = -\frac{(a H)^{D-4}}{D - 4} \; .
\end{eqnarray}
It is also convenient to change variables from $k$ to the time $t_k$
of first horizon crossing for the sum over modes which have experienced
first horizon crossing,
\begin{equation}
k = H(t_k) a(t_k) \qquad \Longrightarrow \qquad \frac{dk}{k} = [1 \!-\! 
\epsilon(t_k)] H(t_k) dt_k \; . \label{dktodt}
\end{equation}
Our final result for $A(t)$ is,
\begin{eqnarray}
\lefteqn{A(t) \simeq -\frac18 \Bigl(\frac{D\!-\!2}{D\!-\!4}\Bigr) 
\frac{[D \!-\! 2 \epsilon(t)] H^{D-2}(t)}{\Gamma(\frac{D-1}{2}) 
(4\pi)^{\frac{D-1}{2}}} - \frac{H^2(t)}{8\pi^2} } \nonumber \\
& & \hspace{5.8cm} + \frac1{4\pi^2} \!\! \int_{t_i}^{t} \!\!\!\! dt' H^3(t') 
\Bigl[1 \!-\! \epsilon(t')\Bigr] C\Bigl( \epsilon(t')\Bigr) \; . \qquad 
\label{Aduring}
\end{eqnarray}
It is worth noting that the divergent first term of (\ref{Aduring}) is
proportional to the Ricci scalar $R(t) = (D-1) [D - 2 \epsilon(t)] H^2(t)$.

\subsection{After Primordial Inflation}

Although expression (\ref{A1}) is generally valid, after the end of inflation
$a(t) H(t)$ falls off and the larger $k$ modes which had experienced first 
horizon crossing undergo a second horizon crossing. If the end of inflation
occurs at $t_e$ then the last mode which experiences first horizon crossing
is $k_e = a(t_e) H(t_e)$ and we must distinguish between ultraviolet modes
($k_e < k < \infty$) which never experienced horizon crossing, intermediate
modes ($a(t) H(t) < k < k_e$) which have undergone both first and second 
crossing, and infrared modes ($k_i < k < a(t) H(t)$) which are still 
super-horizon,
\begin{equation}
\int_{k_i}^{\infty} \!\!\!\! dk = \int_{k_e}^{\infty} \!\!\!\!\!\! dk + 
\int_{Ha}^{k_e} \!\!\!\!\!\! dk + \int_{k_i}^{H a} \!\!\!\!\!\! dk \; . 
\label{intdkafter}
\end{equation}

The ultraviolet integrations are somewhat different from before,
\begin{eqnarray}
\int_{k_e}^{\infty} \!\!\!\! dk \, k^{D-3} & = & \frac{k^{D-2}}{D - 2} 
\Bigl\vert_{k_e}^{\infty} = -\frac{k_e^{D-2}}{D - 2} \longrightarrow 
-\frac12 k_e^2 \; , \\
\int_{k_e}^{\infty} \!\!\!\! dk \, k^{D-5} & = & \frac{k^{D-4}}{D - 4} 
\Bigl\vert_{k_e}^{\infty} = -\frac{k_e^{D-4}}{D - 4} \longrightarrow
-\frac{[a H]^{D-4}}{D - 4} - \ln\Bigl[ \frac{k_e}{a H}\Bigr] \; .
\end{eqnarray}
We convert the intermediate mode sum from $k$ to the time of second crossing,
\begin{eqnarray}
\lefteqn{ \frac1{4\pi^2} \! \int_{aH}^{k_e} \!\! \frac{dk}{k} \,
H^2(t_k) C\Bigl( \epsilon(t_k)\Bigr) \!\times\! \Bigl[ 
\frac{a(t_2(t_k))}{a(t)} \Bigr]^2 \cos^2\Bigl[ \int_{t_2(t_k)}^{t} \!\!\!\!\!
\!\!\!dt'' \frac{k}{a(t'')}\Bigr] } \nonumber \\
& & \hspace{0cm} = \frac1{4\pi^2 a^2(t)} \! \int_{t_e}^{t} \!\!\! dt' 
[\epsilon(t') \!-\! 1] H(t') a^2(t') \cos^2\Bigl[ \int_{t'}^{t} \!\!  
\frac{k dt''}{a(t'')}\Bigr] \!\times\! H^2(t_1) C\Bigl( \!\epsilon(t_1)\!
\Bigr) , \qquad \label{intsimp1} \\ 
& & \hspace{0cm} \simeq \frac1{8\pi^2 a^2(t)} \! \int_{t_e}^{t} \!\!\! dt' 
[\epsilon(t') \!-\! 1] H(t') a^2(t') \!\times\! H^2(t_1) C\Bigl( 
\epsilon(t_1) \Bigr) . \qquad \label{intsimp2}
\end{eqnarray}
The rapid oscillations evident in Figure~\ref{2ndcross} justify replacing
$\cos^2$ by $\frac12$. Super-horizon modes contribute the same as for
(\ref{Aduring}) except that the upper limit is $t_2(t_i)$. Our final result 
for $A(t)$ is,
\begin{eqnarray}
\lefteqn{A(t) \simeq -\frac18 \Bigl(\frac{D\!-\!2}{D\!-\!4}\Bigr) 
\frac{[D \!-\! 2 \epsilon(t)] H^{D-2}(t)}{\Gamma(\frac{D-1}{2}) 
(4\pi)^{\frac{D-1}{2}}} - \frac{[2 \!-\! \epsilon(t)] H^2(t)}{8\pi^2} 
\ln\Bigl[\frac{k_e}{a(t) H(t)}\Bigr] } \nonumber \\
& & \hspace{0cm} - \frac{1}{8\pi^2} \!\times\! \Bigl[ \frac{k_e}{a(t)}
\Bigr]^2 + \frac1{8\pi^2 a^2(t)} \! \int_{t_e}^{t} \!\!\! dt' 
[\epsilon(t') \!-\! 1] H(t') a^2(t') \!\times\! H^2(t_1) C\Bigl( 
\epsilon(t_1) \Bigr) \nonumber \\
& & \hspace{4.5cm} + \frac1{4\pi^2} \!\! \int_{t}^{t_2(t_i)} 
\!\!\!\!\!\!\!\!\!\!\!\! dt' [\epsilon(t') \!-\! 1] H(t') \!\times\! 
H^2(t_1) C\Bigl( \epsilon(t_1)\Bigr) . \qquad \label{Aafter}
\end{eqnarray}

\subsection{General Metric Form}

The expectation value of $\phi^2$ must be a scalar functional of the metric.
We only need it for the class of metrics (\ref{geometry}) in order to model 
cosmology, however, describing changes to the force of gravity requires a 
more general class of metrics. And it is worth mentioning that 1-loop
computations on de Sitter background indicate that the force of gravity can 
experience nonperturbatively strong corrections from inflationary scalars 
\cite{Park:2015kua} and gravitons \cite{Tan:2022xpn}. 

Expressions (\ref{Aduring}) and (\ref{Aafter}) make it plain that the divergence 
in $i\Delta(x;x')$ is proportional to the Ricci scalar. The old result 
(\ref{Sasha1}) of Dolgov and Pellicia \cite{Dolgov:2005se} suggests that we try
to model the finite, nonlocal part as the inverse of the covariant d'Alembertian,
\begin{equation}
\square = -\Bigl( \frac{d}{dt} + 3 H\Bigr) \frac{d}{dt} \; , \label{square}
\end{equation}
acting on some curvature-squared. From the nonlocal part of $A(t)$ during 
inflation (\ref{Aduring}) we find,
\begin{eqnarray}
\lefteqn{\square \! \int_{t_i}^{t} \!\!\!\! dt' H^3 (1 \!-\! \epsilon)
C(\epsilon) = -\Bigl( \frac{d}{dt} + 3 H\Bigr) \Bigl\{ H^3 (1 \!-\! \epsilon)
C(\epsilon)\Bigr\} \; ,} \\
& & \hspace{3cm} = -3 H^4 (1 \!-\! \epsilon)^2 C(\epsilon) + H^3 \Bigl[ 
C(\epsilon) + (1\!-\!\epsilon) C'(\epsilon)\Bigr] \dot{\epsilon} \; . \qquad 
\label{squareA}
\end{eqnarray}

\begin{figure}[H]
\centering
\includegraphics[width=6cm]{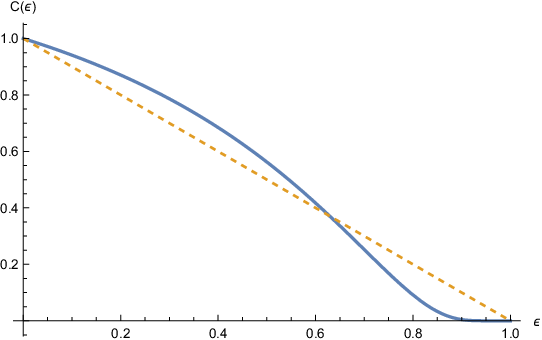}
\caption{\footnotesize Comparison of $C(\epsilon)$ (in solid blue) versus
$1-\epsilon$ (in dashed yellow).} 
\label{Cofepsilon}
\end{figure}

Trying to understand (\ref{squareA}) as something quadratic in the curvature
is challenging because of the complicated function $C(\epsilon)$ given in
expression (\ref{Cdef}). However, it was never realistic to devise a simple
model for the full nonlocal part; what we seek instead is a reasonable
approximation. In this regard it is worth noting that the function 
$C(\epsilon)$ is not far off from $1 - \epsilon$, as shown in 
Figure~\ref{Cofepsilon}.

If we employ $C(\epsilon) \simeq 1 - \epsilon$, and neglect the 
$\dot{\epsilon}$ terms in (\ref{squareA}), it follows that the quadratic 
curvature we seek is approximately,
\begin{equation}
\square \Biggl\{ \frac1{4\pi^2} \! \int_{t_i}^{t} \!\!\!\! dt' H^3 (1 
\!-\! \epsilon) C(\epsilon) \Biggr\} \simeq -\frac3{4\pi^2} \, H^4 (1 \!-\! 
\epsilon)^3 \label{squareA2}
\end{equation} 
Specializing the Riemann tensor and its contractions to the cosmological 
background (\ref{geometry}) gives,
\begin{eqnarray}
R^0_{~i0j} = (1 \!-\! \epsilon) H^2 g_{ij} & , & R^{i}_{~jk\ell} = H^2 \Bigl(
\delta^i_{~k} g_{j\ell} \!-\! \delta^i_{~\ell} g_{jk}\Bigr) , \qquad
\label{Riemann} \\
R_{00} = -3 (1 \!-\! \epsilon) H^2 & , & R_{ij} = (3 \!-\! \epsilon) H^2 
g_{ij} \; . \qquad \label{Ricci}
\end{eqnarray}
Had there been only two factors of $(1 - \epsilon)$, instead of three, we
would have recognized two factors of $R_{00}$
\begin{equation}
\square \Biggl\{ \frac1{4\pi^2} \! \int_{t_i}^{t} \!\!\!\! dt' H^3 (1 
\!-\! \epsilon) C(\epsilon) \Biggr\} \simeq -\frac{R_{00}^2}{12 \pi^2} \; .
\label{squareA3}
\end{equation} 
This is curiously similar to the nonlocal invariant $Y[g]$ which was 
employed \cite{Deffayet:2011sk,Woodard:2014wia,Deffayet:2014lba} to 
construct a relativistic, purely metric realization of Milgrom's Modified 
Newtonian Gravity (MOND) \cite{Milgrom:1983ca,Milgrom:1983pn}. However,
the extra factor of $1 - \epsilon$ spoils this correspondence. Nor does the
$R_{00}^2$ form extend to the post-inflationary regime, during which no new 
modes experience first horizon crossing and the only weak time dependence 
derives from the $1/a^2(t)$ redshift of modes which have undergone second
horizon crossing.
 
\subsection{Instantaneously Constant $\epsilon$ Approximation}

When the first slow roll parameter is constant the amplitude $\alpha(n,\kappa)$ 
turns out to be a simple factor times the norm-squared of a Hankel function of 
the first kind. We define the $\alpha_1$ approximation by replacing the constant
first slow roll parameter of this solution with $\epsilon(n)$,
\begin{equation}
\alpha_1(n,\kappa) \equiv \frac1{2 \kappa e^{(D-2) n}} \!\times\!
\frac{\pi}{2} \!\times\! z(n,\kappa) \!\times\! \Bigl\vert H^{(1)}_{\nu(n)}\Bigl(
z(n,\kappa) \Bigr) \Bigr\vert^2 \equiv \frac{\gamma_1(n,\kappa)}{2 \kappa 
e^{(D-2) n}} \; . \label{alpha1}
\end{equation}
Here the argument $z(n,\kappa)$ and the index $\nu(n)$ are,
\begin{equation}
z(n,\kappa) \equiv \frac{\kappa e^{-n}}{(1 \!-\! \epsilon) \chi} = 
\frac{\zeta(n,\kappa)}{1 \!-\! \epsilon} \qquad , \qquad \nu(n) \equiv 
\frac{(D \!-\! 1 \!-\! \epsilon)}{2 (1 \!-\! \epsilon)} . \label{znudef}
\end{equation}
The large $z$ asymptotic form of the Hankel function implies,
\begin{equation}
\gamma_1 = \frac{\pi z}{2} \Bigl\vert H^{(1)}_{\nu}(z) \Bigr\vert^2 = 1 + 
\frac{(\nu^2 \!-\! \frac14)}{2 z^2} + \frac{3 (\nu^2 \!-\! \frac14) 
(\nu^2 \!-\! \frac94)}{8 z^4} + O(z^{-6}) \; . \label{Hankelexp}
\end{equation}
From the definitions (\ref{znudef}) of $z(n,\kappa)$ and $\nu(n)$ we infer
the relations,
\begin{eqnarray}
\frac{(\nu^2 \!-\! \frac14)}{z^2} &\!\!\! = \!\!\!& \frac{(D\!-\!2) 
(D \!-\! 2 \epsilon)}{4 \zeta^2} \; , \qquad \label{term1} \\
\frac{(\nu^2 \!-\! \frac94)}{z^2} &\!\!\! = \!\!\!& \frac{(D\!-\!4\!+\! 2\epsilon) 
(D \!+\! 2 \!-\! 4\epsilon)}{4 \zeta^2} \; . \qquad \label{term2}
\end{eqnarray}
It follows that the large $z$ expansion of $\gamma_1$ gives all the terms
on the first line of (\ref{gammaexp}),
\begin{equation}
\gamma_1 = 1 \!+\! \frac{(D\!-\!2) (D\!-\!2\epsilon)}{8 \zeta^2} \!+\! 
\frac{3 (D\!-\!2) (D \!-\! 2 \epsilon) (D \!-\! 4 \!+\! 2 \epsilon) (D \!+\! 2
\!-\! 4 \epsilon)}{128 \zeta^4} + O(\zeta^{-6}) \; . \label{gamma1exp}
\end{equation}
However, the ultraviolet expansion of $\gamma_1(n,\kappa)$ fails to recover
the terms involving derivatives of $\epsilon$ on the second line of
(\ref{gammaexp}). Therefore, the instantaneously constant $\epsilon$
approximation reproduces the ultraviolet divergences of $A(t)$ and $B_{\mu}(t)$,
but it misses some divergences in $C_0(t)$ and $\overline{C}(t)$.

The instantaneously constant $\epsilon$ approximation also gives misleading
results for the infrared. To see this we evaluate the coincidence limit of the
propagator in this approximation \cite{Janssen:2008px},
\begin{equation}
i\Delta_1(x;x) = \frac{[(1 \!-\! \epsilon) H]^{D-2}}{(4\pi)^{\frac{D}2}} 
\frac{\Gamma(2 \!-\! \frac{D}2)}{\frac{D}2 \!-\! 1} \frac{\Gamma(\frac{D-1}{2}
\!+\! \nu) \Gamma(\frac{D-1}{2} \!-\! \nu)}{\Gamma(\frac12 \!+\! \nu) 
\Gamma(\frac12 \!-\! \nu)} \; . \label{Delta1}
\end{equation}
The complicated ratio of Gamma functions in (\ref{Delta1}) can be expanded 
around $\delta \equiv 4 - D$,
\begin{eqnarray}
\lefteqn{\frac{\Gamma(\frac{D-1}{2} \!+\! \nu) \Gamma(\frac{D-1}{2} \!-\! \nu)}{
\Gamma(\frac12 \!+\! \nu) \Gamma(\frac12 \!-\! \nu)} = \Bigl[ \Bigl(\frac{D \!-\! 3}{2}
\Bigr)^2 \!-\! \nu^2\Bigr] \!\times\! \frac{\Gamma(\frac{1}{2} \!+\! \nu \!-\! 
\frac{\delta}{2}) \Gamma(\frac{1}{2} \!-\! \nu) \!-\! \frac{\delta}{2})}{
\Gamma(\frac12 \!+\! \nu) \Gamma(\frac12 \!-\! \nu)} \; ,} \\
& & \hspace{-0.5cm} = -\frac{(D\!-\!2) (2 \!-\! \epsilon) [2 \!+\! (D\!-\!4) \epsilon]}{
4 (1 \!-\! \epsilon)^2} \Biggl\{ 1 \!-\! \psi\Bigl(\frac12 \!+\! \nu\Bigr) 
\frac{\delta}{2} \!-\! \psi\Bigl(\frac12 \!-\! \nu\Bigr) \frac{\delta}{2} + O(\delta^2) 
\Biggr\} . \qquad \label{ratio}
\end{eqnarray}
Note the unphysical poles in the finite parts,
\begin{eqnarray}
\epsilon = 1 - \frac1{N} & \Longrightarrow & \psi\Bigl(\frac12 - \nu\Bigr) 
\longrightarrow \pm \infty \; , \qquad \label{poleslow} \\
\epsilon = 1 + \frac1{N+1} & \Longrightarrow & \psi\Bigl(\frac12 + \nu\Bigr) 
\longrightarrow \pm \infty \; . \qquad \label{poleshigh}
\end{eqnarray}
These poles arise because the instantaneously constant $\epsilon$ mode sum 
contains infrared divergences for all values in the range $0 \leq \epsilon \leq 
\frac32$ \cite{Ford:1977in}. However, in most cases the divergences are of the 
power-law type which dimensional regularization automatically subtracts. For 
the special values (\ref{poleslow}-\ref{poleshigh}) one of the infrared 
divergences happens to become logarithmic, at which point dimensional 
regularization registers it as a divergence at $D=4$ \cite{Janssen:2008px}.
 
The infrared divergences (\ref{poleslow}-\ref{poleshigh}) are unphysical 
for two reasons. First, the actual amplitude freezes in to a constant 
(\ref{freezein}) which depends on the first slow roll parameter at horizon
crossing, $\epsilon(n_k)$, rather than the evolving value $\epsilon(n)$. 
Second, modes which were already super-horizon at the beginning of inflation 
($k < k_i$) would not have experienced the evolution necessary to reach their 
freeze-in amplitudes (\ref{freezein}). These small modes can either be 
assumed to exist in some less infrared singular state \cite{Vilenkin:1983xp}, 
or else they can be discarded altogether, which would pertain if the spatial 
manifold were compact \cite{Tsamis:1993ub}.

Although the infrared divergences (\ref{poleslow}-\ref{poleshigh}) can be 
repaired by changing the mode sum \cite{Janssen:2008px}, their presence
indicates a profound problem with the instantaneously constant $\epsilon$
approximation. It seems best to avoid the approximation altogether, in spite
of the fact that it {\it was} employed in similar studies of coincident 
inflationary propagators \cite{Kyriazis:2019xgj,Sivasankaran:2020dzp,
Katuwal:2021kry}. The key difference between those cases and this one is 
that nonzero masses protected the mode sum from infrared divergences.

\section{Determining $B_{\mu}(t)$, $C_0(t)$ and $\overline{C}(t)$}

The purpose of this section is to use approximations (\ref{Aduring}) and
(\ref{Aafter}) for $A(t)$ to infer approximations for $B_{\mu}(t)$, $C_0(t)$
and $\overline{C}(t)$. We begin by discussing the generic procedure for 
reconstructing $C_0(t)$ and $\overline{C}(t)$ from $A(t)$. Then approximate
results are derived for $B_{\mu}(t)$, $C_0(t)$ and $\overline{C}(t)$, first 
during inflation and then after the end of inflation.

\subsection{Generic Considerations for $C_0(t)$ and $\overline{C}(t)$}

The Introduction defined two linear combinations of $C_0$ and $\overline{C}$, 
\begin{equation}
\Biggl\{ \begin{matrix}
C_A \equiv& -C_0 + (D\!-\!1) \overline{C} \\
C_B \equiv& C_0 + \overline{C}
\end{matrix} \Biggr\} \quad \Longleftrightarrow \quad \Biggl\{
\begin{matrix}
C_0 =& -\frac1{D} C_A + (\frac{D-1}{D}) C_B \\
\overline{C} =& \frac1{D} C_A + \frac1{D} C_B
\end{matrix} \Biggr\} . \label{CmnfromAB} 
\end{equation}
Relation (\ref{Sasha1}) gives $C_A$ in terms of $\square A$,
\begin{equation}
C_A(t) = -\frac12 \Bigl[ \ddot{A}(t) + (D\!-\!1) H(t) \dot{A}(t)\Bigr] \; .
\label{CAfromA}
\end{equation}
The other linear combination derives from conservation of the scalar
stress tensor (\ref{Sasha4}),
\begin{equation}
2 \Bigl[ \frac{d}{dt} + D H \Bigr] C_B = -\Bigl( 
\frac{D\!-\!2}{D\!-\!1}\Bigr) \dot{C}_A = \frac12 \Bigl(
\frac{D\!-\!2}{D\!-\!1}\Bigr) \Bigl[ \dddot{A} + (D\!-\!1) H \ddot{A}
- (D\!-\!1) \epsilon H^2 \dot{A}\Bigr] \; . \label{CBfromA}
\end{equation}

It is desirable to extract a factor of $(\frac{d}{dt} + D H)$ from
the right hand side of (\ref{CBfromA}) as much as possible,
\begin{equation}
\dddot{A} + (D\!-\!1) H \ddot{A} - (D\!-\!1) \epsilon H^2 \dot{A} =
\Bigl[\frac{d}{dt} + D H\Bigr] \Bigl[\ddot{A} - H \dot{A} \Bigr] 
+ D (1 \!-\! \epsilon) H^2 \dot{A} \; . \label{extract}
\end{equation}
Hence we can write,
\begin{eqnarray}
\lefteqn{C_B(t) = \frac14 \Bigl( \frac{D\!-\!2}{D\!-\!1}\Bigr) \Bigl[
\ddot{A}(t) - H(t) \dot{A}(t)\Bigr] } \nonumber \\
& & \hspace{3.5cm} + \frac{D (D\!-\!2)}{4 (D\!-\!1) a^D(t)} \int_{t_i}^{t} 
\!\!\!\! dt' \, a^D(t') H^2(t') [1 \!-\! \epsilon(t')] \dot{A}(t') \; . 
\qquad \label{CBintform}
\end{eqnarray}
A special case is the divergent part of $A(t)$, which we see from 
(\ref{Aduring}) and (\ref{Aafter}) takes the form of $[D - 2 \epsilon]
H^{D-2}$. For this dependence, the final term of (\ref{extract}) allows
another factor of $(\frac{d}{dt} + D H)$ to be extracted, up to a
remainder proportional to $(D-4)$,
\begin{eqnarray}
\lefteqn{D (1\!-\!\epsilon) H^2 \frac{d}{dt} \Bigl[ (D \!-\! 2 \epsilon) 
H^{D-2}\Bigr] } \nonumber \\
& & \hspace{0.5cm} = \Bigl[ \frac{d}{dt} + D H\Bigr] \Bigl[-(D\!-\!2) \epsilon 
(D\!-\!2 \epsilon) H^D\Bigr] + (D\!-\!4) \epsilon' (D \!-\! 2 \epsilon) 
H^{D+1} \; . \qquad \label{extractdiv} 
\end{eqnarray}
Relation (\ref{extractdiv}) guarantees that the divergent parts of $C_0(t)$ 
and $\overline{C}(t)$ are local,
\begin{eqnarray}
{C_0}_{\rm div} & \!\!\! = \!\!\! & \frac14 \Bigl\{ \ddot{A}_{\rm div} + H
\dot{A}_{\rm div} - \frac{(D\!-\!2)^2}{D} \, \epsilon H^2 A_{\rm div} \Bigr\} 
\; , \qquad \label{C0div} \\
{\overline{C}}_{\rm div} & \!\!\! = \!\!\! & -\frac1{4 (D\!-\! 1)} \Bigl\{
\ddot{A}_{\rm div} + (2 D \!-\! 3) H \dot{A}_{\rm div} + \frac{(D\!-\!2)^2}{D}
\, \epsilon H^2 A_{\rm div} \Bigr\} \; . \qquad \label{Cbardiv}
\end{eqnarray}

The divergence structure (\ref{C0div}-\ref{Cbardiv}) is consistent with an
$R^2$ counterterm,
\begin{equation}
\Delta \mathcal{L} = \alpha R^2 \sqrt{-g} \qquad \Longrightarrow \qquad
\Delta T_{\mu\nu} = 2 \alpha \Bigl[ R_{\mu\nu} - \frac14 g_{\mu\nu} R +
g_{\mu\nu} \square - D_{\mu} D_{\nu} \Bigr] R \; . \label{DTmn}
\end{equation}
The induced change in $C_{\mu\nu}$ is,
\begin{equation}
\Delta C_{\mu\nu} = \Delta T_{\mu\nu} - \frac{g_{\mu\nu} g^{\rho\sigma} 
T_{\rho\sigma}}{D \!-\! 2} = 2 \alpha \Bigl[ R_{\mu\nu} - \frac{g_{\mu\nu} 
R}{2 (D\!-\!2)} - \frac{g_{\mu\nu} \square}{D \!-\! 2} - D_{\mu} D_{\nu} 
\Bigr] R \; . \label{DCmn}
\end{equation}
Using the $D$-dimensional relations $R_{00} = -(D-1) (1 - \epsilon) H^2$
and $R_{ij} = (D-1-\epsilon) H^2 g_{ij}$, we find the time-time and 
space-space components,
\begin{eqnarray}
\Delta C_{00} & \!\!\! = \!\!\! & -2 \alpha \Bigl( \frac{D\!-\!1}{D\!-\!2}
\Bigr) \Bigl[ \Bigl( \frac{D \!-\! 4}{2}\Bigr) H^2 - (D\!-\!3) \epsilon H^2 + 
\frac{d^2}{dt^2} + H \frac{d}{dt} \Bigr] R \; , \qquad \label{DC00} \\
\Delta \overline{C} & \!\!\! = \!\!\! & \frac{2 \alpha}{D \!-\! 2} \Bigl[
\frac{(D \!-\! 1) (D\!-\! 4)}{2} H^2 + \epsilon H^2 + \frac{d^2}{dt^2} 
+ (2 D \!-\! 3) H \frac{d}{dt} \Bigr] R \; . \qquad \label{DCbar}
\end{eqnarray}
Comparison between expressions (\ref{C0div}-\ref{Cbardiv}) and 
(\ref{DC00}-\ref{DCbar}) implies that the divergences can be canceled by,
\begin{equation}
\alpha = -\frac1{64} \Bigl( \frac{D\!-\!2}{D\!-\!1}\Bigr)^2 
\frac1{\Gamma(\frac{D-1}{2}) (4\pi)^{\frac{D-1}{2}}} \times 
\frac{\mu^{D-4}}{D\!-\!4} \; , \label{alpha}
\end{equation}
where $\mu$ is the mass scale of dimensional regularization.

\subsection{During Primordial Inflation}

From expression (\ref{Aduring}) we can identify divergent and finite 
parts of $A(t)$ during inflation,
\begin{eqnarray}
A_{\rm div} & \!\!\! = \!\!\! & -\frac18 \Bigl(\frac{D\!-\!2}{D\!-\!4}\Bigr) 
\frac{[D \!-\! 2 \epsilon(t)] H^{D-2}(t)}{\Gamma(\frac{D-1}{2}) 
(4\pi)^{\frac{D-1}{2}}} \; , \label{Adiv} \\
A_{\rm fin} & \!\!\! = \!\!\! & - \frac{H^2(t)}{8\pi^2} + \frac1{4\pi^2} \!\! 
\int_{t_i}^{t} \!\!\!\! dt' H^3(t') \Bigl[1 \!-\! \epsilon(t')\Bigr] 
C\Bigl( \epsilon(t')\Bigr) \; . \label{Afin}
\end{eqnarray}
An important derived quantity is the finite residual at the end of
relation (\ref{extractdiv}),
\begin{equation}
\lim_{D \rightarrow 4} (D\!-\!4) \epsilon' H^3 A_{\rm div} = 
-\frac{H^5}{8 \pi^2} (2 \!-\! \epsilon) \epsilon' \; . \label{residual}
\end{equation}
Another key quantity is the first time derivative of the finite part,
\begin{equation}
\dot{A}_{\rm fin} = \frac{H^3}{4 \pi^2} - \frac{H^3}{4 \pi^2} (1 \!-\!
\epsilon) \Bigl( 1 \!-\! C(\epsilon) \Bigr) \; . \label{Afindot}
\end{equation}
Our result for $B_{\mu}(t)$ involves (\ref{Adiv}) and (\ref{Afindot}),
\begin{equation}
B_{\mu} = \frac12 \Bigl\{\dot{A}_{\rm div} + \dot{A}_{\rm fin}\Bigr\}
\delta^0_{~\mu} \; . \label{Bduring}
\end{equation}
Taking $\epsilon = 0$ in (\ref{Bduring}) agrees with the known de Sitter
limit (\ref{dSprop'}).

We have already derived results (\ref{C0div})-\ref{Cbardiv}) for the
divergent parts of $C_0(t)$ and $\overline{C}(t)$. To get the finite
parts we must first derive the finite parts of $C_A(t)$ and $C_B(t)$.
The first of these is,
\begin{equation}
{C_{A}}_{\rm fin} = -\frac12 \Bigl( \frac{d}{dt} + 3 H\Bigr) 
\dot{A}_{\rm fin} \; . \label{CAfin}
\end{equation}
To get the finite part of $C_B$ it is best to extract a factor of 
$(\frac{d}{dt} + 4 H)$ from $4 (1 - \epsilon) H^2$ times the first term
of (\ref{Afindot}),
\begin{equation}
4 (1 \!-\! \epsilon) H^2 \dot{A}_{\rm fin} = \Bigl( \frac{d}{dt} + 4 H
\Bigr) \Bigl\{ \frac{H^4}{4 \pi^2} \Bigr\} - \frac{H^5}{8 \pi^2} \!\times\!
8 ( 1 \!-\! \epsilon)^3 [1 \!-\! C(\epsilon)] \; . \label{extract2}
\end{equation}
It follows that the finite part of $C_B$ is,
\begin{eqnarray}
\lefteqn{{ C_B}_{\rm fin} = \frac16 \Bigl( \frac{d}{dt} \!-\! H\Bigr) 
\dot{A}_{\rm fin} + \frac{H^4}{24 \pi^2} } \nonumber \\
& & \hspace{2cm} - \Bigl[ \frac{d}{dt} \!+\! 4 H\Bigr]^{-1} \Biggl\{
\frac{H^5}{48 \pi^2} \Bigl[ 8 (1 \!-\! \epsilon)^3 [1 \!-\! C(\epsilon)] +
(2 \!-\! \epsilon) \epsilon'\Bigr] \Biggr\} . \qquad \label{CBfin}
\end{eqnarray}

Inserting expressions (\ref{CAfin}) and (\ref{CBfin}) into (\ref{CmnfromAB})
gives,
\begin{eqnarray}
C_0 & \!\!\! = \!\!\! & {C_0}_{\rm div} + \frac14 \Bigl( \frac{d}{dt} \!+\!
H\Bigr) \dot{A}_{\rm fin} + \frac{H^4}{32 \pi^2} \nonumber \\
& & \hspace{1.5cm} - \Bigl[ \frac{d}{dt} \!+\! 4 H\Bigr]^{-1} \Biggl\{
\frac{H^5}{64 \pi^2} \Bigl[ 8 (1 \!-\! \epsilon)^3 [1 \!-\! C(\epsilon)] +
(2 \!-\! \epsilon) \epsilon'\Bigr] \Biggr\} , \qquad \label{C0during1} \\
\overline{C} & \!\!\! = \!\!\! & \overline{C}_{\rm div} - \frac1{12} \Bigl( 
\frac{d}{dt} \!+\! 5 H\Bigr) \dot{A}_{\rm fin} + \frac{H^4}{96 \pi^2} 
\nonumber \\
& & \hspace{1.5cm} - \Bigl[ \frac{d}{dt} \!+\! 4 H\Bigr]^{-1} \Biggl\{
\frac{H^5}{192 \pi^2} \Bigl[ 8 (1 \!-\! \epsilon)^3 [1 \!-\! C(\epsilon)] +
(2 \!-\! \epsilon) \epsilon'\Bigr] \Biggr\} . \qquad \label{Cbarduring1} 
\end{eqnarray}
Recall that the divergent parts were given in expressions (\ref{C0div}) and
(\ref{Cbardiv}). 

It is straightforward to check that taking $\epsilon = 0$ in 
(\ref{C0during1}-\ref{Cbarduring1}) agrees with the known de Sitter limit 
(\ref{dSprop''}). It is also worth noting that the inverse differential
operators on the last lines of (\ref{C0during1}-\ref{Cbarduring1}) become
effectively local during inflation,
\begin{equation}
\Bigl[ \frac{d}{dt} \!+\! 4 H\Bigr]^{-1} \Bigl[H^5 f(\epsilon)\Bigr](t) 
\equiv \frac1{a^4(t)} \! \int_{t_i}^{t} \!\!\! dt' a^4(t') H^5(t') 
f\Bigl(\epsilon(t')\Bigr) \simeq \frac{H^4(t) f(\epsilon(t))}{4 [1\!-\! 
\epsilon(t)]} \; . \label{local}
\end{equation}
We can therefore write,
\begin{eqnarray}
C_0 & \!\!\! \simeq \!\!\! & {C_0}_{\rm div} + \frac14 \Bigl( \frac{d}{dt} 
\!+\! H\Bigr) \dot{A}_{\rm fin} + \frac{H^4}{32 \pi^2} \Bigl[ 1 - (1 \!-\!
\epsilon)^2 [1 \!-\! C(\epsilon)] - \frac{(2 \!-\! \epsilon) \epsilon'}{8
(1 \!-\! \epsilon)} \Bigr] , \qquad \label{C0during} \\
\overline{C} & \!\!\! \simeq \!\!\! & \overline{C}_{\rm div} - \frac1{12} 
\Bigl( \frac{d}{dt} \!+\! 5 H\Bigr) \dot{A}_{\rm fin} + \frac{H^4}{96 \pi^2}
\Bigl[ 1 - (1 \!-\! \epsilon)^2 [1 \!-\! C(\epsilon)] - \frac{(2 \!-\! 
\epsilon) \epsilon'}{8 (1 \!-\! \epsilon)} \Bigr] . \qquad 
\label{Cbarduring}
\end{eqnarray} 

\subsection{After Primordial Inflation}

The divergent part (\ref{Adiv}) is unchanged after the end of inflation,
whereas the finite part becomes, 
\begin{eqnarray}
\lefteqn{A_{\rm fin} = \frac{(2 \!-\! \epsilon) H^2}{8 \pi^2} \ln\Bigl(
\frac{k_e}{a H}\Bigr) - \frac{H^2}{8 \pi^2} \Bigl( \frac{k_e}{a H}\Bigr)^2 
+ \frac1{4\pi^2 a^2(t)} \! \int_{t_e}^{t} \!\!\! dt' 
[\epsilon(t') \!-\! 1] H(t') a^2(t') } \nonumber \\
& & \hspace{0cm} \times `` \tfrac12 {\rm "} \!\times\! H^2(t_1) C\Bigl( 
\epsilon(t_1) \Bigr) + \frac1{4\pi^2} \!\! \int_{t}^{t_2(t_i)} 
\!\!\!\!\!\!\!\!\!\!\!\! dt' [\epsilon(t') \!-\! 1] H(t') \!\times\! 
H^2(t_1) C\Bigl( \epsilon(t_1)\Bigr) \; . \qquad \label{Afinafter}
\end{eqnarray}
Here $`` \tfrac12 {\rm "}$ indicates the factor of $\cos^2[\int_{t_2}^{t} dt'
k/a(t')]$ in expression (\ref{A1}). Although this factor averages to $\tfrac12$
inside the integral, it gives unity when evaluated at $t '= t$. This means that 
the time derivative receives no contribution from acting on the upper and lower 
limits of the two integrals in (\ref{Afinafter}),
\begin{eqnarray}
\lefteqn{ \dot{A}_{\rm fin} = -\frac{[2 \epsilon (2 \!-\! \epsilon) +
\epsilon'] H^3}{8 \pi^2} \ln\Bigl(\frac{k_e}{a H}\Bigr) - \frac{(1 \!-\! \epsilon)
(2 \!-\! \epsilon) H^3}{8 \pi^2} + \frac{H^3}{4 \pi^2} \Bigl( \frac{k_e}{a H}
\Bigr)^2 } \nonumber \\
& & \hspace{2.5cm} - \frac{H(t)}{4 \pi^2 a^2(t)} \!\! \int_{t_e}^{t} \!\!\! dt' 
[\epsilon(t') \!-\! 1] H(t') a^2(t') \!\times\! H^2(t_1) C\Bigl( \epsilon(t_1)
\Bigr) . \qquad \label{Afindotafter}
\end{eqnarray}

It turns out that the last line of (\ref{Afindotafter}) provides the dominant 
time dependence. To see this, note first that the factor of $H(t') a^2(t')$ inside 
the integral is constant during radiation domination. Because the factor of 
$H^2(t_1) C(\epsilon(t_1))$ is nearly constant, it follows that the integral 
grows nearly linearly during radiation domination. During radiation domination 
the multiplicative factor of $H(t)/a^2(t)$ falls off like inverse time-squared,
which makes the entire term fall off like inverse time. The last line of
(\ref{Afindotafter}) also represents the nonlocal memory effect associated with 
the universe having undergone primordial inflation. Had radiation domination 
($\epsilon = D/2$ in dimensional regularization) pertained for all time both 
$A_{\rm div}$ and $A_{\rm fin}$ would have vanished.

The quantity $B_{\mu}(t)$ takes the same form (\ref{Bduring}) after inflation 
as during,
\begin{equation}
B_{\mu} = \frac12 \Bigl\{\dot{A}_{\rm div} + \dot{A}_{\rm fin}\Bigr\}
\delta^0_{~\mu} \; . \label{Bafter}
\end{equation}
However, one must use the post-inflationary expression (\ref{Afindotafter})
for $\dot{A}_{\rm fin}(t)$, even though $A_{\rm div}(t)$ is unchanged from
(\ref{Adiv}). Note that $B_{\mu}(t)$ falls off like inverse time-squared
during radiation-domination, instead of vanishing as it would without having
undergone primordial inflation.

Because the divergent part of $A(t)$ is the same (\ref{Adiv}) as during 
inflation, the divergent parts of $C_0(t)$ and $\overline{C}(t)$ are unchanged
from expressions (\ref{C0div}-\ref{Cbardiv}). The finite parts of $C_A(t)$
and $C_B(t)$ are,
\begin{eqnarray}
{C_A}_{\rm fin} & \!\!\!\! = \!\!\!\! & -\frac12 \Bigl( \frac{d}{d t} \!+\! 
3 H\Bigr) \dot{A}_{\rm fin} \; , \qquad \label{CAfinafter} \\
{C_B}_{\rm fin} & \!\!\!\! = \!\!\!\! & \frac16 \Bigl( \frac{d}{d t} \!-\! 
H\Bigr) \dot{A}_{\rm fin} + \Bigl[ \frac{d}{dt} \!+\! 4 H\Bigr]^{-1} \Biggl\{ 
\frac23 (1 \!-\! \epsilon) H^2 \dot{A}_{\rm fin} \!-\! \frac{H^5}{48\pi^2} 
(2 \!-\! \epsilon) \epsilon' \Biggr\} . \qquad \label{CBfinafter}
\end{eqnarray}
Substituting in (\ref{CBfromA}) gives,
\begin{eqnarray}
\lefteqn{C_0 = {C_0}_{\rm div} + \frac14 \Bigl( \frac{d}{d t} \!+\! H\Bigr) 
\dot{A}_{\rm fin} } \nonumber \\
& & \hspace{3.5cm} + \Bigl[ \frac{d}{dt} \!+\! 4 H\Bigr]^{-1} \!\Biggl\{ 
\frac12 (1 \!-\! \epsilon) H^2 \dot{A}_{\rm fin} \!-\! \frac{H^5}{64\pi^2} 
(2 \!-\! \epsilon) \epsilon' \Biggr\} , \qquad \label{C0after} \\
\lefteqn{\overline{C} = \overline{C}_{\rm div} - \frac1{12} \Bigl( 
\frac{d}{d t} \!+\! 5 H\Bigr) \dot{A}_{\rm fin} } \nonumber \\
& & \hspace{3.5cm} + \Bigl[ \frac{d}{dt} \!+\! 4 H\Bigr]^{-1} \!\Biggl\{ 
\frac16 (1 \!-\! \epsilon) H^2 \dot{A}_{\rm fin} \!-\! \frac{H^5}{192\pi^2} 
(2 \!-\! \epsilon) \epsilon' \Biggr\} . \qquad \label{Cbarafter}
\end{eqnarray}
Recall that $\dot{A}_{\rm fin}$ is the post-inflationary expression 
(\ref{Afindotafter}) while the divergent parts are unchanged from
(\ref{C0div}-\ref{Cbardiv}).

During radiation-domination the term inside the curly brackets of 
expressions (\ref{C0after}-\ref{Cbarafter}) falls off like $1/t^3$. The 
inverse differential operator acting on it is,
\begin{equation}
\Bigl[ \frac{d}{dt} \!+\! 4 H\Bigr]^{-1} f(t) = \frac1{a^4(t)} \!
\int_{t_i}^{t} \!\!\! dt' a^4(t') f(t') \; . \label{inversedef}
\end{equation}
Hence the integral grows logarithmically, and the entire term falls off
like $\ln(t)/t^2$. This is marginally stronger than the finite terms on 
the first lines of (\ref{C0after}-\ref{Cbarafter}). Note again the 
effect of the universe having undergone primordial inflation; $C_{\mu\nu}(t)$
vanishes for eternal radiation-domination.

\section{Conclusions}

We have studied the massless, minimally coupled scalar (\ref{MMCSL}) 
on a general cosmological background (\ref{geometry}) which undergoes 
primordial inflation. Our goal was to derive good analytic approximations 
for the coincidence limits of the scalar propagator and its first two 
derivatives (\ref{threeDs}) which we call $A(t)$, $B_{\mu}(t)$ and 
$C_{\mu\nu}(t)$. Because dynamical graviton and scalar modes obey the 
same equation \cite{Lifshitz:1945du}, these correlators are crucial to 
building models of the quantum gravitational back-reaction on inflation 
\cite{Tsamis:1997rk}, and to {\it deriving} these models by re-summing 
the large logarithms induced by loops of inflationary gravitons 
\cite{Miao:2021gic}. Each of our analytic approximations was tested
numerically using a plausible expansion history developed in section 2.3.
However, we stress that the analytic approximations are valid for {\it 
any} geometry (\ref{geometry}) which experiences primordial inflation,
and do not depend upon the parameters introduced in section 2.3.

Our strategy was to derive $B_{\mu}(t)$ and $C_{\mu\nu}(t)$ from $A(t)$. 
In section 3 we expressed $A(t)$ as a dimensionally regulated, spatial 
Fouri\-er mode sum (\ref{Asum}) of an amplitude $\mathcal{A}(t,k)$. In 
the ultraviolet this amplitude has a series expansion, (\ref{gammazeta}) 
and (\ref{gammaexp}), in powers of the small parameter $(aH/k)^2$. 
Figure~\ref{hc10} shows that just the first two terms of this expansion 
provide an excellent approximation until the moment of first horizon 
crossing $k = a(t_k) H(t_k)$. After first horizon crossing, and before 
second crossing, the amplitude freezes in to a constant value 
(\ref{freezein}-\ref{Cdef}). Figures~\ref{hc10} and \ref{2ndcross} 
demonstrate that this approximation remains valid until the mode 
experiences second horizon crossing after the end of inflation. At this
point Figure~\ref{2ndcross} shows that the amplitude is well 
approximated by a damped oscillatory form (\ref{post2}).

Section 4 employed our approximations for the amplitude $\mathcal{A}(t,k)$ 
to estimate the mode sum (\ref{Asum}) for $A(t)$. Because the damped 
oscillatory form (\ref{post2}) only pertains after the end of inflation, 
we derived separate approximations for $A(t)$ during inflation 
(\ref{Aduring}) and afterwards (\ref{Aafter}). In each case these 
approximations consist of the same local divergent part (\ref{Adiv}) plus 
a nonlocal finite part whose form depends upon whether or not primordial 
inflation has ended. 

The coincident first derivative $B_{\mu}(t)$ always takes the form
(\ref{Bduring}) in terms of the $\dot{A}_{\rm div}$ and $\dot{A}_{\rm fin}$.
During inflation the finite part can be approximated by (\ref{Afindot}),
whereas our post-inflationary approximation is (\ref{Afindotafter}). The 
coincident second derivative has two distinct components (\ref{Sasha3}), 
whose divergent parts are (\ref{C0div}-\ref{Cbardiv}). Their finite parts 
are (\ref{C0during}-\ref{Cbarduring}) during inflation and 
(\ref{C0after}-\ref{Cbarafter}) afterwards.

Two major conclusions of our work concerning all three post-inflationary 
correlators are (1) that they show the effect of the universe having 
undergone primordial inflation and (2) that they transmit the high scales
of primordial inflation to late times. To see the first point, note that
radiation domination corresponds to $\epsilon = \frac2{D}$ in dimensional 
regularization. From expression (\ref{znudef}) we see that this takes the 
index $\nu$ of the constant epsilon mode function to be $\nu = -\frac12$, 
which is indistinguishable from its flat space value of $\nu = +\frac12$.
Hence the coincidence limit of the constant epsilon propagator (\ref{Delta1}) 
vanishes in dimensional regularization, as do the coincidence limits of all 
derivatives. This vanishing is apparent in the divergent part (\ref{Adiv}) 
of our result for the actual correlator, and in the first term of its
finite part (\ref{Afinafter}). However, the two integrals in expression
(\ref{Afinafter}) for $A_{\rm fin}(t)$ reach back to the time of primordial 
inflation and carry inflationary expansion rates to the epoch of radiation 
domination,
\begin{eqnarray}
\frac1{8\pi^2 a^2(t)} \! \int_{t_e}^{t} \!\!\! dt' [\epsilon(t') \!-\! 1] 
H(t') a^2(t') \!\times\! H^2(t_1) C\Bigl( \epsilon(t_1) \Bigr) \simeq 
\frac{H^2_{\rm inf}}{8\pi^2} \; , \qquad \label{Int1} \\
\frac1{4\pi^2} \!\! \int_{t}^{t_2(t_i)} \!\!\!\!\!\!\!\!\!\!\!\! dt' 
[\epsilon(t') \!-\! 1] H(t') \!\times\! H^2(t_1) C\Bigl( \epsilon(t_1)\Bigr) 
\simeq \frac{\ln[\frac{t_2(t_i)}{t}] H^2_{\rm inf}}{4\pi^2} \; . \qquad 
\label{Int2}
\end{eqnarray}
The same comments pertain as well to $B_{\mu}(t)$ and $C_{\mu\nu}(t)$.

One thing we have not be able to do is represent $A(t)$ using a simple, 
nonlocal scalar. The approximations we derived, (\ref{Aduring}) and 
(\ref{Aafter}), are only valid for a general cosmological geometry 
(\ref{geometry}). This should suffice for constructing modified gravity 
models of cosmology \cite{Tsamis:1997rk,Deser:2007jk,Woodard:2014iga,
Deser:2019lmm}, but it does not extend to models whose purpose is to 
alter the gravitational force \cite{Soussa:2003vv,Deffayet:2011sk,
Woodard:2014wia,Deffayet:2014lba}. During inflation we showed that $A(t)$
is nearly,
\begin{equation}
t < t_e \qquad \Longrightarrow \qquad A \simeq -\frac1{\square} 
\Biggl[\frac{R^2_{00}}{12 \pi^2} \Biggr] . \label{Aapprox}
\end{equation}
However, this relation is not quite right, and there is no comparably
simple form after inflation.

Our method has been to derive results for the {\it primitive} correlators
(\ref{threeDs}), without committing to any specific renormalization 
condition. With these results in hand, one can understand the paradoxical 
sign that Dolgov and Pelliccia found for the $\ln(a)$ term in the de 
Sitter limit of $A(t)$ \cite{Dolgov:2005se}, which disagrees with the 
famous early result (\ref{olddS}) \cite{Vilenkin:1982wt,Linde:1982uu,
Starobinsky:1982ee}, and with our dimensionally regulated result 
(\ref{dSprop}) \cite{Onemli:2002hr,Onemli:2004mb}. Dolgov and Pelliccia 
followed the inverse of our procedure, computing $C_{\mu\nu}(t)$ and then 
using relation (\ref{Sasha1}) to reconstruct $A(t)$. There is nothing 
wrong with this. Nor is there anything wrong with them expressing 
$C_{\mu\nu}$ in terms of the scalar stress tensor,
\begin{equation} 
C_{\mu\nu} = T_{\mu\nu} - \frac1{D \!-\! 2} g_{\mu\nu} g^{\rho\sigma} 
T_{\rho\sigma} \; . \label{stress}
\end{equation}
The problem is that they chose to renormalize the scalar stress tensor
with the counterterm (\ref{DTmn}), rather than isolating $A_{\rm div}$ as 
we did. Renormalizing $T_{\mu\nu}$, rather than $A$, leads to some finite,
contributions from expressions (\ref{DC00}-\ref{DCbar}) which do not 
vanish for de Sitter. It is those finite residuals which change the sign 
of the $\ln(a)$ part of $A(t)$ in expressions (\ref{olddS}) and 
(\ref{dSprop}). Although the two renormalization schemes differ only by
finite, local changes at the level of the correlator $C_{\mu\nu}$, the 
difference in $A(t)$ is not local, so the scheme Dolgov and Pelliccia
must be rejected.

\vskip 1cm

\centerline{\bf Acknowledgements}

We are grateful for discussions with A. D. Dolgov and N. C. Tsamis.
MU was supported by a scholarship from Consejo Nacional de Ciencia 
y Tecnología (CONACYT) of Mexico. RPW was supported by NSF grants 
PHY-1912484 and PHY-2207514, and by the Institute for Fundamental 
Theory at the University of Florida.

\end{document}